%% file: aa56978-25.tex
\documentclass{aa}  
\usepackage{float}
\usepackage{booktabs}
\usepackage{multirow}
\usepackage{siunitx}
\usepackage[colorlinks,linkcolor={blue},citecolor={blue},urlcolor={blue}]{hyperref} 
\usepackage{graphicx}
\usepackage{subfigure}
\usepackage{txfonts}   
\usepackage{comment}
\usepackage{color}
\usepackage{xcolor}
\colorlet{avi}{green!0!orange!100!}

\begin{document} 

\titlerunning{Influence of variable IMF on the dynamical black hole mass estimates in massive ETGs}
\authorrunning{A. Chaturvedi et al.}
   \title{Influence of variable initial mass function on the dynamical black hole mass estimates in massive early-type galaxies}
   \author{Avinash Chaturvedi  \inst{1, 2}
          \and
          Sabine Thater\inst{3}\
          \and 
          Davor Krajnovi\'{c}\inst{2}\
          \and
          Ignacio Mart\'{i}n-Navarro\inst{4}
          \and
          Alina Boecker\inst{3}\
           }

   \institute{Finnish Centre for Astronomy with ESO (FINCA),
        University of Turku, 20014 Turku, Finland\\
            \email{avinash.chaturvedi@utu.fi}
    \and 
    Leibniz-Institute for Astrophysics Potsdam (AIP), An der Sternwarte 16, D-14482 Potsdam, Germany
    \and
    Department of Astrophysics, University of Vienna, T\"urkenschanzstraße 17, 1180 Vienna, Austria
    \and
    Instituto de Astrofísica de Canarias, E-38200 La Laguna, Tenerife, Spain}

   \date{Received: 25 August 2025/ Accepted: 17 November, 2026}

%\date{Received September 15, 1996; accepted March 16, 1997}

% \abstract{}{}{}{}{} 
% 5 {} token are mandatory
 
  \abstract
   {Several studies have shown that the stellar initial mass function (IMF) in massive early-type galaxies (ETGs) is not universal. Recent results from stellar population synthesis models, based on integral-field unit (IFU) observations, indicate that the IMF varies radially, becoming increasingly bottom-heavy towards galaxy centres and implying an excess of low-mass dwarf stars. This radial variation alters the stellar mass-to-light ratio ($\rm M_{*}/L$) and can significantly affect dynamical black hole (BH) mass estimates in massive ETGs. In this study, we investigate the impact of a radially varying IMF on BH mass estimates for five massive galaxies: NGC3923, NGC4261, NGC4636, IC4296, and IC4329. Using DYNAMITE, a triaxial Schwarzschild orbit-based modelling code, we perform mass modelling of these galaxies based on MUSE and SINFONI IFU observations. Our models incorporate radially varying stellar $\rm M_{*}/L$ profiles derived from spatially resolved stellar population modelling with a free IMF, and we compare the resulting BH masses with those obtained from dynamical models assuming a constant stellar $\rm M_{*}/L$. All galaxies in our sample show bottom-heavy IMFs in their central regions, with the IMF generally becoming lighter towards larger radii. We obtain constrained BH masses for NGC3923 and IC4329, while only upper limits are obtained for NGC4261, NGC4636, and IC4296. Compared with the corresponding constant-$\rm M_{*}/L$ dynamical models, the variable-$\rm M_{*}/L$ models produce galaxy-dependent changes in the inferred BH masses and upper-limit constraints. For NGC3923 and IC4329, the inferred BH masses decrease by approximately 83\% and 11\%, respectively, while for the galaxies with only upper limits the constraints can either become higher or lower.  Our results demonstrate that spatial variations in the IMF and stellar $\rm M_{*}/L$ can have a non-negligible but strongly galaxy-dependent impact on dynamical BH mass measurements. These results highlight the importance of accounting for such variations in future dynamical modelling efforts.}


   \maketitle
%
%-------------------------------------------------------------------

\section{Introduction}

Measurements of the different mass components in massive early-type galaxies (ETGs)—notably black hole (BH), stellar component, and dark matter (DM) —provide critical constraints on galaxy formation models. These components exhibit tight scaling relationships with one another, indicating that they co-evolve alongside the growth and evolution of the galaxy. The BH mass strongly correlates with various galaxy properties such as velocity dispersion, bulge mass, and luminosity, suggesting a close link between the evolution of BHs and galaxies \citep{Ferrarese2005, Kormendy2013, VandenBosch2016, Saglia2016}. Accurate measurements of these components are crucial for understanding these relationships, and by extension, for refining galaxy formation models. However, a major challenge arises from the degeneracy of these components in their contribution to the total mass, making it difficult to disentangle and accurately decompose the total mass into its individual components.

The knowledge about the stellar mass and other stellar properties of ETGs can be derived with the stellar population models \citep{Tinsley1980, Trager2000, Thomas2005, Conroy2012} whereas the BH mass and total mass can be derived from the dynamical modelling studies. Multiple studies of massive galaxies based on stellar population modelling either using the line strength or the full spectrum fitting techniques, have shown spatial variations in properties such as the mass-to-light ratio (M/L), color, metallicity, age, alpha-element abundance ratio, and the initial mass function (IMF) \citep{Worthey1992, Mehlert2003, Sanchez2006, Spolaor2009, Kuntschner2010, Conroy2012, Sanchez2014,  Gonzalez2015, Smith2020}. These variations in stellar properties essentially reflect the formation and evolutionary histories of massive galaxies. Massive galaxies are believed to be evolved in two evolutionary phases: first in-situ formation with an intense burst of star formation and second ex-situ with the growth in outer envelopes resulting from the infall or merger events of low mass galaxies \citep{Oser2010, Dokkum2010}.

One of the key, yet debated, findings from stellar population models of massive galaxies is the existence of a variable initial mass function (IMF) in ETGs \citep[see also \citealt{Cappellari2012} for IMF variation from a dynamical modelling perspective]{VanDokkum2010, Spiniello2011, VanDokkum2012, Ferreras2013, LaBarbera2013, Navarrojun2015, Spiniello2014, Conroy2014, Parikh2018}. Although the IMF was previously thought to be invariant based on studies of the Milky Way, most stellar population studies of ETGs have found a Salpeter-like or even steeper IMF slope in their central regions, suggesting an excess of low-mass dwarf stars in the cores of massive galaxies. With advances in IFU data and techniques, recent studies such as \cite{Navarro2019, Navarro2021, Navarro2024} have demonstrated that this IMF variability is confined to the inner few kiloparsecs and may be correlated with the metallicity of the galaxies. More recently, \cite{Parikh2024} have confirmed IMF variability using a sample of seven massive galaxies based on MUSE IFU data.

In contrast to the directly measured stellar M/L from the SSP models, information on the BH mass comes from the dynamical mass modelling studies or, in active galactic nuclei, from methods such as reverberation mapping, single-epoch virial estimates, and megamaser measurements. Most BH mass estimates have been derived using dynamical modelling of stellar or gas kinematics. Although an indirect approach, dynamical modelling can also be used to infer the IMF of galaxies. If the DM halo contributions are properly accounted for and subtracted from the total mass, then the dynamical stellar M/L ratio can be determined. The ratio of this dynamical stellar $\rm M/L_{dyn}$ to the reference stellar $\rm M/L_{ssp}$, obtained from the stellar population models assuming a specific IMF, known as the excess mass factor, can be used to infer the potential change or variation in the IMF. Based on this approach, similar to the results from stellar population modelling, several dynamical modelling studies of massive galaxies have suggested the need for a variable IMF in ETGs \citep{Treu2010, Auger2010, Thomas2011, Dutton2012, Cappellari2012, Sonnenfeld2019, Smith2020, Poci2022, Mehrgan2024, DenBrok2024}. 

Studying a large sample of stacked ETGs from the SDSS data, \cite{Faber2010} suggested that a variable IMF or DM halo is necessary to explain the observed trends in the dynamical $\rm M/L_{dyn}$ ratio. Using orbit-based modelling including the DM halo, \cite{Thomas2011} reached a similar conclusion, finding that the IMF varies with galaxy velocity dispersion, becoming Salpeter-like as the velocity dispersion increases. In the same vein, an analysis of 260 galaxies from the ATLAS3D survey \cite{Cappellari2012, Cappellari2013} found that, on a galaxy-by-galaxy basis, the IMF correlates with galaxy velocity dispersion, transitioning from Kroupa to Salpeter as velocity dispersion rises. However, other studies based on gravitational lensing have favoured a Kroupa-like mass function \citep{Smith2015}.

Although reaching a definitive consensus on IMF variation in massive ETGs remains challenging, most studies support a Salpeter-like or even steeper IMF slope, particularly in their central regions. The results from SSP modelling, which reveal spatial variations in stellar properties—including IMF variation—strongly emphasize the need to account for and assess the impact of these variations on BH mass measurements in massive ETGs. \cite{Bernardi2018} demonstrated that neglecting IMF variation in mass modelling can lead to an overestimation of dynamical stellar $\rm M_{*}/L$ values, which can, in turn, affect the inferred BH mass.

Most previous dynamical modelling studies of galaxies have assumed a constant stellar $\rm M_{*}/L$ ratio when deriving BH and total galaxy mass. While easy to implement, this assumption is unrealistic and simplifies the mass modelling process. Several studies have explored the impact of a variable $\rm M_{*}/L$ in dynamical modelling, revealing degeneracies between the measured BH mass and the $\rm M_{*}/L$ ratio \citep{Lasker2013, McConnel2016, Nguyen2017, Poci2019, Thater2019, Thaterfeb2022, Mehrgan2024}.  More recently, \cite{Thater2023} investigated the effects of incorporating both a variable $\rm M_{*}/L$ with a free IMF in the mass modelling of nucleated ETG FCC47. They found that, unlike a constant $\rm M_{*}/L$, a variable $\rm M_{*}/L$—whether from a fixed or variable IMF—can lower the BH mass by upto $\sim$ 50\%. Extending this line of research is critical to properly accounting for the uncertainty in BH mass measurements arising from dynamical modelling.

This study highlights the impact of including a variable IMF on supermassive black hole (SMBH) mass measurements derived from triaxial orbit-based dynamical modelling. Using the sample of galaxies from \citealt[][(hereafter Thater2026)]{Thater2026} we investigate the impact of variable IMF on the measured BH masses of five massive galaxies—NGC 4261, NGC 4636, NGC 3923, IC 4296 and IC 4329. Thater2026 measured the BH masses of these galaxies by modelling their SINFONI and MUSE kinematics with the orbit based DYNAMITE Schwarzschild modelling approach. To achieve our goal, we adopt similar kinematics, stellar mass models and dynamical modelling method used in Thater2026, except that we incorporate a radially varying stellar mass-to-light $\rm M_{*}/L$ ratio driven by free IMF obtained through the stellar population modelling into the stellar mass models. We then perform the Schwarzschild orbit-based modelling to derive the BH masses of these galaxies for a radially varying stellar $\rm M_{*}/L$ driven by a free IMF. 

The structure of this paper is as follows: In Section \ref{Sec2:Data}, we provide a brief overview of the data and kinematics of our sample galaxies. Section \ref{Sec3:FIF_model} outlines the methodology used for the stellar population analysis. In Section \ref{Sec4:Dynamical_model}, we describe the dynamical modelling approach employed in this study. Section \ref{Sec5:Results} presents the results of both the stellar population and dynamical modelling. In Section \ref{Sec6:Discussion}, we compare and discuss our findings in the context of previous studies. Finally, in Section \ref{Sec7:Summary}, we summarize our conclusions.

%--------------------------------------------------------------------
\section{Data and Kinematics}\label{Sec2:Data}

In this paper, we use a set of five massive galaxies from the SMASHING sample presented in Thater2026 that have IFU observations in both the SINFONI \citep{Eisenhauer2003, Bonnet2004} and MUSE \citep{Bacon2010} spectrographs. A detailed description of the observations and kinematic extraction from SINFONI and MUSE can be found in Thater2026. Here we briefly summarise the main points of the observations and kinematics extraction.

\subsection{Kinematics extraction}
For kinematic extraction in both SINFONI and MUSE observations, we used the Python-implemented penalized pixel fitting method \citep[pPXF;][]{Cappellari2004, Cappellari2017}. pPXF performs full-spectrum fitting of the observed spectrum by convolving a set of template stars with a parametric line-of-sight velocity distribution (LOSVD) modelled as a Gaussian-Hermite series \citep{Gerhard1993, Vandermarel1993}.
Before running pPXF on the galaxy spectrum, we performed Voronoi tessellation \citep{Cappellari2003} on the SINFONI cubes to obtain an optimal, high signal-to-noise (SN) spectrum. We used a target SN ratio between 50 and 70 for Voronoi binning, ensuring that the inner spaxels remain unbinned while maintaining high resolution in the outer regions. For the SINFONI observations, we used the Gemini spectral library of near-IR late-type stars \citep{Winge2009} and a spectral fitting range between 2.20 and 2.40 $\rm \mu$m.

In the case of the MUSE observations, we performed Voronoi tessellation with a target SN of 50 for all galaxies except NGC 4636 and obtained about 2000-2500 spatial bins for each galaxy. For NGC 4636 we used a target SN of 150 to reduce the number of spatial bins. For the stellar templates in the pPXF run, we used the MILES stellar templates \citep{2006MNRAS.371..703S} and chose a spectral range between 4650- 6850 $\AA$. We limited the spectral fitting range at 6850 $\AA$ to avoid the spectral regions affected by telluric features.  

%FIGURE 1
\begin{figure*}
    \centering
        \includegraphics[width=0.90\linewidth]{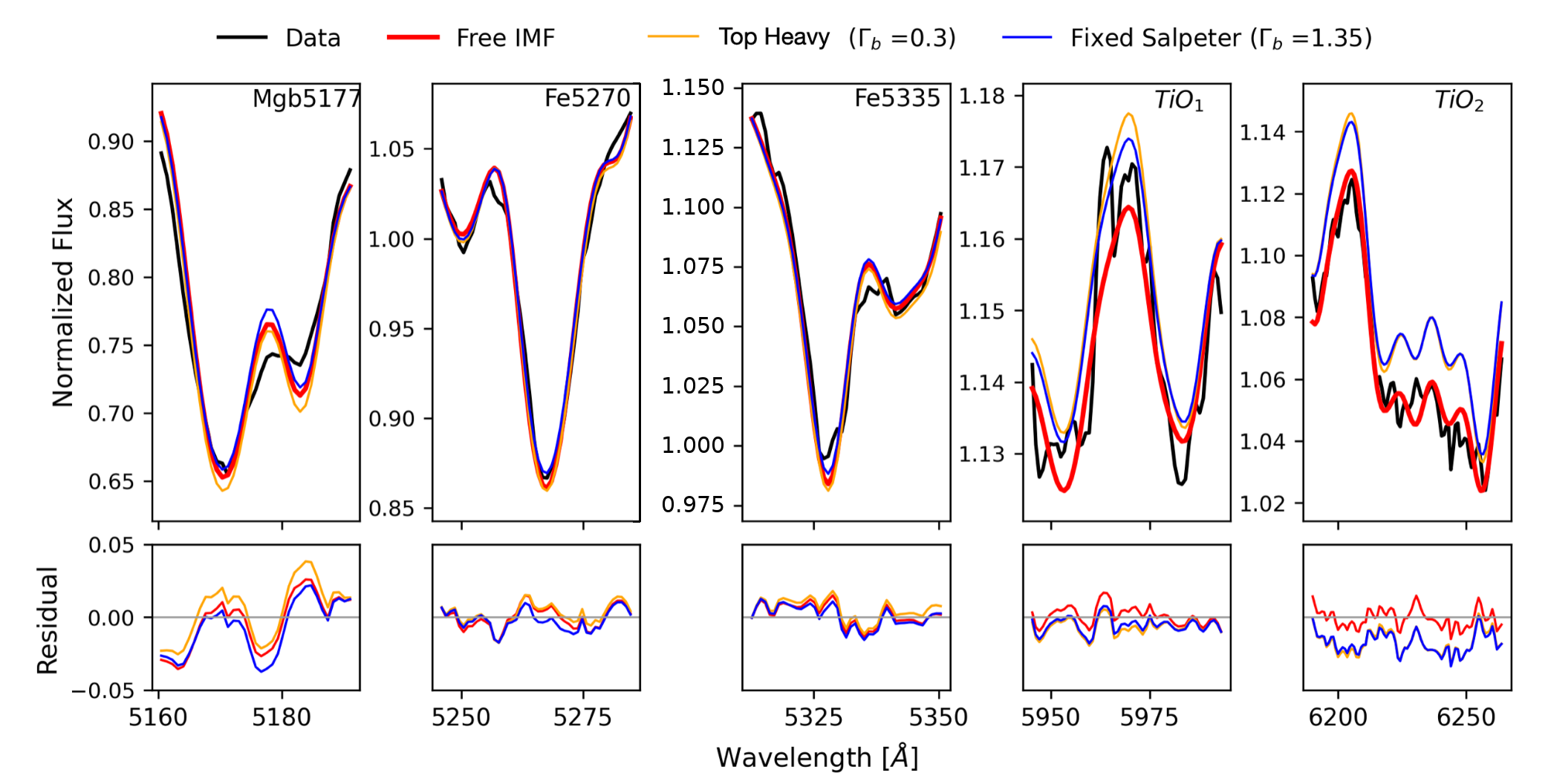}
    \caption{Example of the full-index-fitting analysis for a spatial bin within 5 arcsec of the centre of NGC~4636. Top row: in each panel, the black curve shows the continuum-corrected observed spectrum, while the red, orange, and blue curves show the best-fitting free-IMF model and the models obtained by fixing the IMF to Kroupa-like and Salpeter slopes, respectively. Bottom row: residuals between the observed spectrum and each corresponding model.}
    \label{fig:exm_fifline}
\end{figure*}

\section{Stellar population modelling: Full Index fitting method} \label{Sec3:FIF_model}

To obtain a stellar population model for galaxies, it is crucial to precisely measure variations in simple stellar population (SSP) features that are sensitive to the IMF, age, and metallicity. We used the same spatial bins from the MUSE data, previously employed for kinematic extraction, to perform the stellar population modelling.

To investigate the stellar population properties of our sample galaxies, we applied the full index fitting (FIF) method, as presented in \cite{Navarro2019, Navarro2021}. FIF functions as a hybrid approach between individual line indices and full-spectrum fitting. Instead of fitting the entire spectral range, FIF models individual absorption line features in pixel space within a defined bandpass, after removing the continuum shape using a linear fit to the blue and red continuum regions.
By fitting only specific absorption features, FIF significantly reduces the number of free parameters in the modelling process, thereby decreasing computational time. Additionally, compared to traditional line analysis, FIF increases the number of independent observables, improving the robustness of the results.

%FIGURE 2
\begin{figure}
        \includegraphics[width=1.0\linewidth]{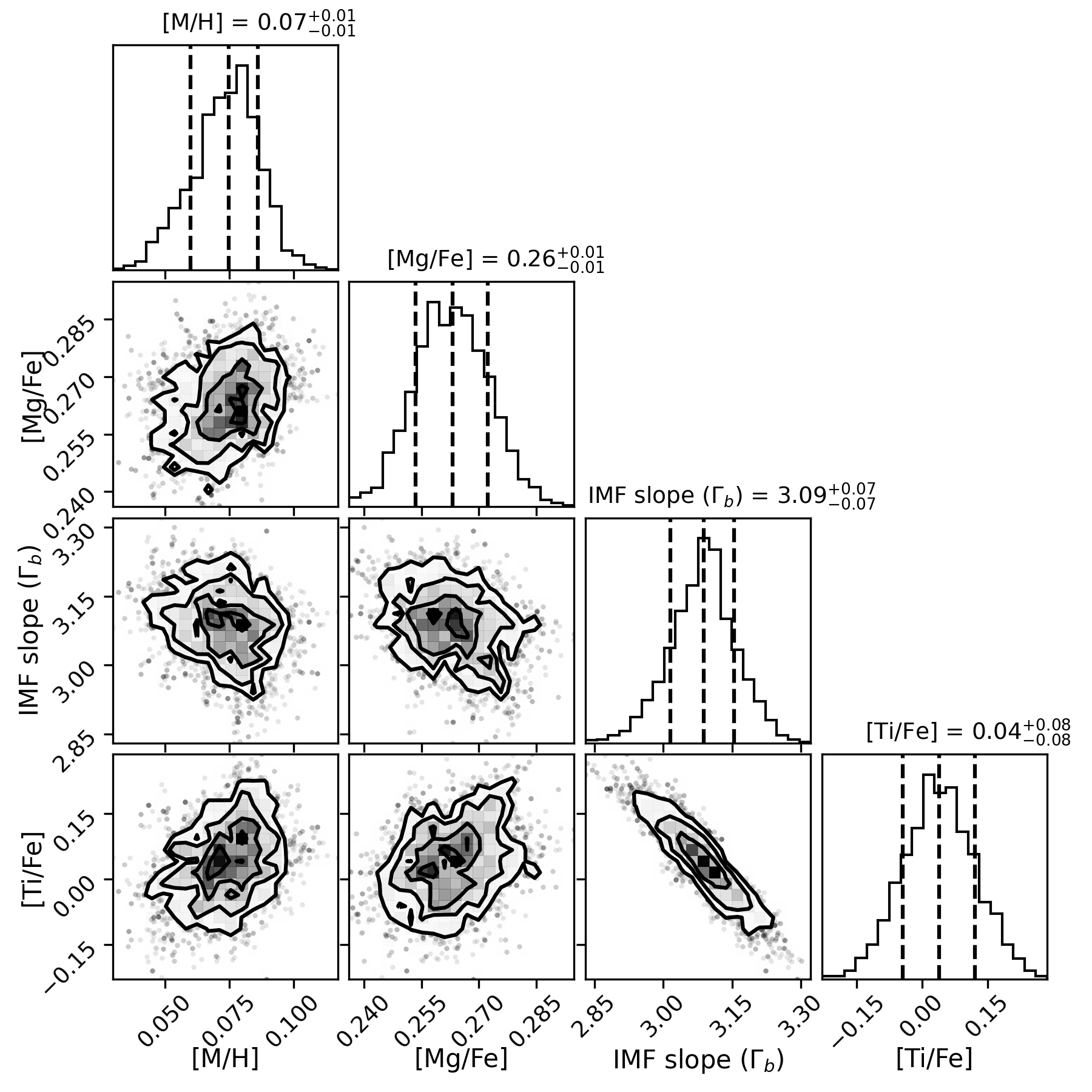}
    \caption{Example corner plot showing the posterior distributions and best-fit values of the FIF parameters for the same spatial bin of NGC 4636 as shown in Figure~\ref{fig:exm_fifline}.}
    \label{fig:exm_fifcorner}
\end{figure}

%FIGURE 3
\begin{figure*}
    \centering
        \includegraphics[width=0.95\linewidth]{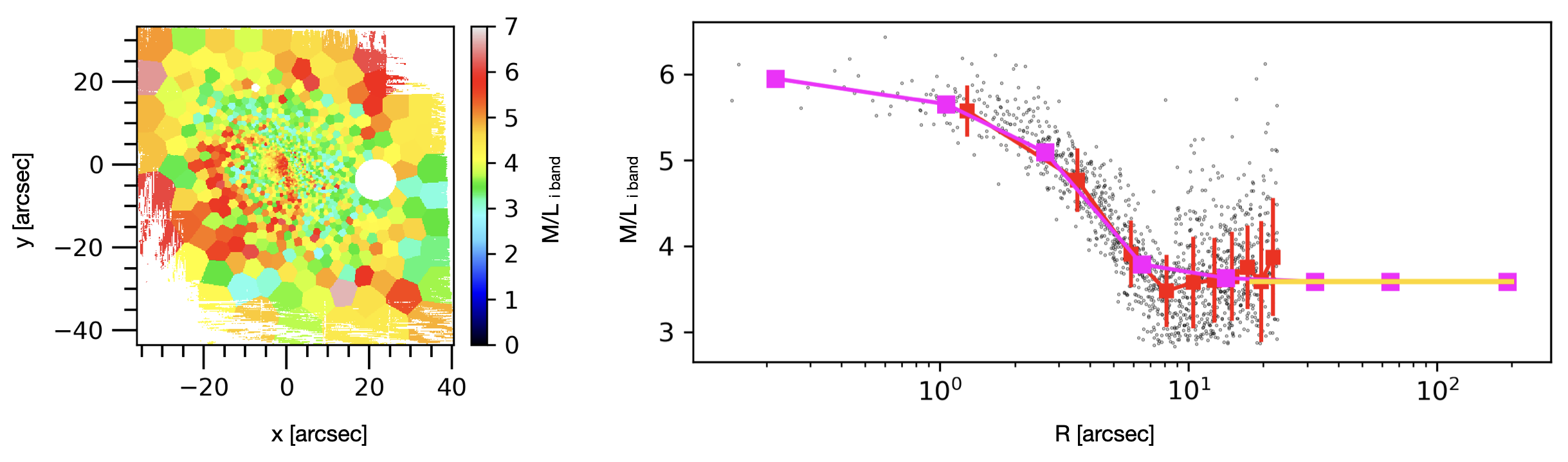}
    \caption{Stellar $\rm M_{*}/L$ map of NGC3923 (left panel) and the interpolated and extrapolated radial $\rm M_{*}/L$ values of NGC3923 (right panel), with extrapolation applied at radii $>20$ arcsec. Red squares denote the radially binned $M_{*}/L$ values obtained from the FIF analysis. Grey dots denote the individual spatial bins of the stellar $\rm M_{*}/L$ map. Magenta squares show the interpolated or extrapolated $\rm M_{*}/L$ values evaluated on the same radial grid as the MGE models. Stellar $\rm M_{*}/L$ maps for the other galaxies are shown in Appendix~\ref{fig:Appendix_mge_mlmap}.}
    \label{fig:NGC4636_ml_scaled}
\end{figure*}

We follow a similar methodology to previous work based on the FIF implementation presented in \cite{Navarro2019, Navarro2021, Navarro2024} to derive spatial maps of stellar population properties and the IMF.  Here we briefly summarise the main components and steps of FIF modelling. Measuring IMF variation with FIF involves three main steps. First, we apply the pPXF method to the Voronoi spatial bins to measure kinematics, including only the line-of-sight mean velocity and stellar velocity dispersion, without attempting to constrain stellar population parameters. In the second step, we re-run pPXF using the kinematic parameters derived in the first step as fixed inputs and perform a regularised fit over the age–metallicity–IMF slope parameter space using the MILES stellar population models to obtain luminosity-weighted stellar ages. Fixing the kinematics helps mitigate degeneracies between stellar kinematics and population properties \citep{Sanchez2011}. In the second step, age and metallicity are determined simultaneously, and the resulting age is used as a fixed input parameter for the subsequent FIF final step. By fitting age and metallicity together under regularisation, the age–metallicity degeneracy is mitigated in a standard way, following previous works \citep{Sanchez2011, Navarro2019}.
In the third step FIF method refines the stellar population parameters by fitting individual absorption features sensitive to metallicity [M/H], abundance ratios ([Mg/Fe], [Ti/Fe]) and the IMF slopes.

For both pPXF and FIF analysis, we utilize the MILES alpha-variable stellar population models presented in \citet{Vazdekis2015}. These models use the BaSTI isochrones \citep{Pietrinferni2004, Pietrinferni2006}, calculated at [$\rm \alpha/Fe$] = 0.4 and at the solar scale ($\rm \alpha/Fe$ = 0.0). The MILES models cover a metallicity range from $-2.25$ to 0.26 and ages from 0.03 to 14 Gyr. For the IMF functional form, these models adopt the bimodal shape \citep{Vazdekis1996}, parametrized through the logarithmic high-mass-end slope ($\rm \Gamma_{b}$). This bimodal IMF parametrization controls the high-mass-end slope, while at the same time effectively varying the relative number of low-mass stars, as the integral of the IMF is normalized to $1\,\rm M_{\odot}$. However, for the old stellar populations in our sample ($\sim10$--14 Gyr), the main-sequence turn-off mass is approximately $0.8$--$1.0\,\rm M_{\odot}$, and therefore our analysis does not directly constrain the original high-mass end of the IMF. For reference, values of $\rm \Gamma_{b}=1.3$ and 1.35 correspond to Milky Way Kroupa-like \citep{Kroupa2002} and Salpeter-like \citep{Salpeter1955} mass functions, respectively.

We selected five absorption line features: Fe5270, Fe5335, Mgb 5177, TiO1 and TiO2 and fit for four stellar population parameters. For NGC~3923 and IC~4329, the TiO2 feature is strongly affected by residual skyline contamination and was therefore excluded from the FIF analysis. For these two galaxies, the fit was performed using Fe5270, Fe5335, Mgb~5177, and TiO1. To derive the best fit parameters, FIF utilises the emcee Bayesian Markov chain Monte Carlo sampler \citep{Emcee2013}. Figure \ref{fig:exm_fifline} shows an example of the FIF fit for a central spatial bin within 5 arcsec of NGC 4636. In addition to the best-fitting free-IMF model, we show models with fixed Kroupa-like and Salpeter IMF slopes to illustrate the sensitivity of the fitted spectral features to the assumed IMF. The corresponding posterior distributions of the fitted parameters, assuming a free IMF, are shown in Figure \ref{fig:exm_fifcorner}. We used the best-fit parameter values derived from FIF to obtain the variable stellar mass-to-light ratio ${\rm M_{*}/L}$ in the sdss-i band. To do this, we matched the best-fit FIF parameters to the corresponding photometric predictions from the $\alpha$-variable MILES SSP models based on BaSTI isochrones, adopting the nearest grid point to obtain the SDSS $i$-band magnitude.

\section{Dynamical mass-modelling}\label{Sec4:Dynamical_model}

We employed the orbit-based Schwarzschild dynamical modelling technique \citep{Schwarzschild1979} for our mass modelling work, a method that has been robustly developed and tested across various systems, with its first triaxial implementation carried out by \cite{Vandenbosch2008}. Our goal is to incorporate a variable stellar mass-to-light ($\rm M_{*}/L$) ratio into our dynamical modelling, which we obtain from the FIF analysis (Section \ref{Sec3:FIF_model}) using a variable IMF. Orbit-based dynamical method requires generating a gravitational potential for the galaxy to produce model luminosity and kinematics that match real observations. This potential consists of three main components: the BH, stellar mass, and the dark matter (DM) halo. In the following subsection, we describe the stellar mass models derived using the Multi-Gaussian Expansion (MGE)\footnote{https://pypi.org/project/mgefit/} method \citep{Emsellem1994, Cappellari2002} and the setup of our mass modelling.

\subsection{Stellar mass modelling}\label{Sec4:Stellar_mass_model}

For the stellar mass models, we utilize the MGE approach to parametrize the surface brightness of the galaxies. The MGE method fits a set of two-dimensional Gaussians to the observed galaxy images, which can be analytically deprojected into a three-dimensional density distribution. To construct an accurate stellar mass model, the MGE approach requires a combination of high-resolution and large spatial extent images. For this purpose, we used both HST and ground-based photometric images available for each galaxy in our sample. We adopt the MGE values of our sample galaxies from Thater2026 (see their Section 5.1, Figure 3 and Table 3 of the Supplementary Material).

Now to incorporate a spatially varying mass-to-light ratio ($\rm M_{*}/L$) into the stellar mass models, we use the $\rm M_{*}/L$ map derived from our full index fitting (FIF) analysis (Section \ref{Sec3:FIF_model}).  First, we radially bin the $\rm M_{*}/L$ values and interpolate them to match the same radial grid as the one produced by the MGE model. We then multiply the interpolated $\rm M_{*}/L$ values by the MGE-derived surface brightness to produce the stellar mass map, which accounts for the variable $\rm M_{*}/L$. Since the stellar MGE models extend beyond the MUSE data for the galaxies, we need to extrapolate the $\rm M_{*}/L$ in the outskirts to match the radial extent of MGE models. For this purpose, we calculate the median $\rm M_{}/L$ from the outermost reliable radial bins, typically between 15 and 20 arcsec, where the S/N remains sufficiently high, and adopt this value as a constant beyond the MUSE coverage. The left panel of Figure~\ref{fig:NGC4636_ml_scaled} shows an example of the stellar $\rm M_{*}/L$ map obtained from the FIF analysis. The right panel shows the radially binned $\rm M_{*}/L$ measurements from the FIF map (red squares) and the corresponding values interpolated onto the MGE radial grid (magenta squares). The outer magenta squares, connected by the solid yellow line, indicate the constant $\rm M_{*}/L$ values adopted in the extrapolated region.

%FIGURE 5
\begin{figure*}
    \centering
        \includegraphics[width=1.0\linewidth]{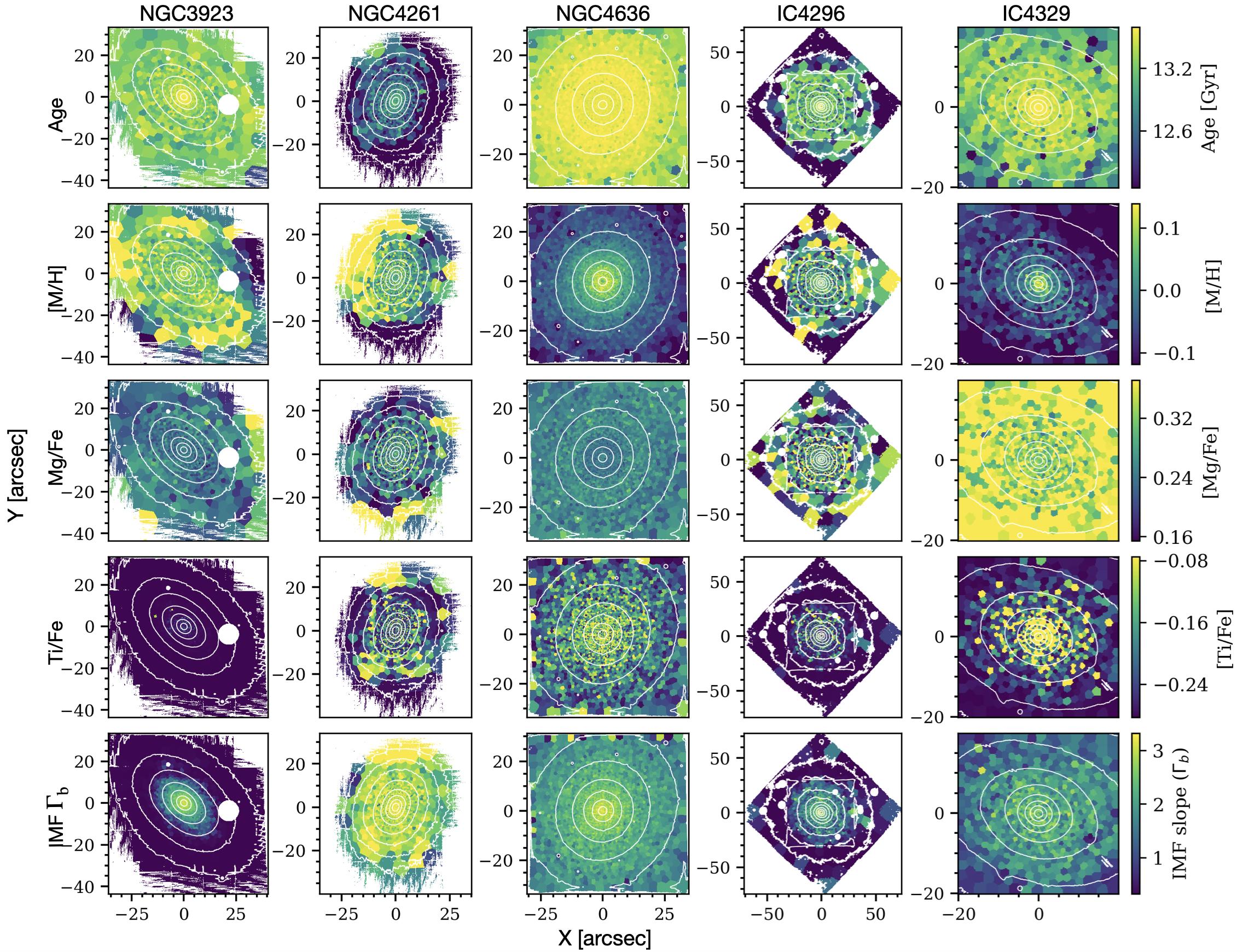}
    \caption{2D stellar population maps of our sample galaxies obtained from FIF analysis. From top to bottom, each row presents maps of stellar age, metallicity [M/H], [Mg/Fe], [Ti/Fe], and IMF slope ($\rm \Gamma_{b})$, respectively, for each galaxy (columns), as labelled in the top panels. Overplotted white colour contours indicate the isophotes of the MUSE white-light image of galaxies.}
    \label{fig:2Dfif_maps}
\end{figure*}

%FIGURE 5
\begin{figure}
        \includegraphics[width=1.0\linewidth]{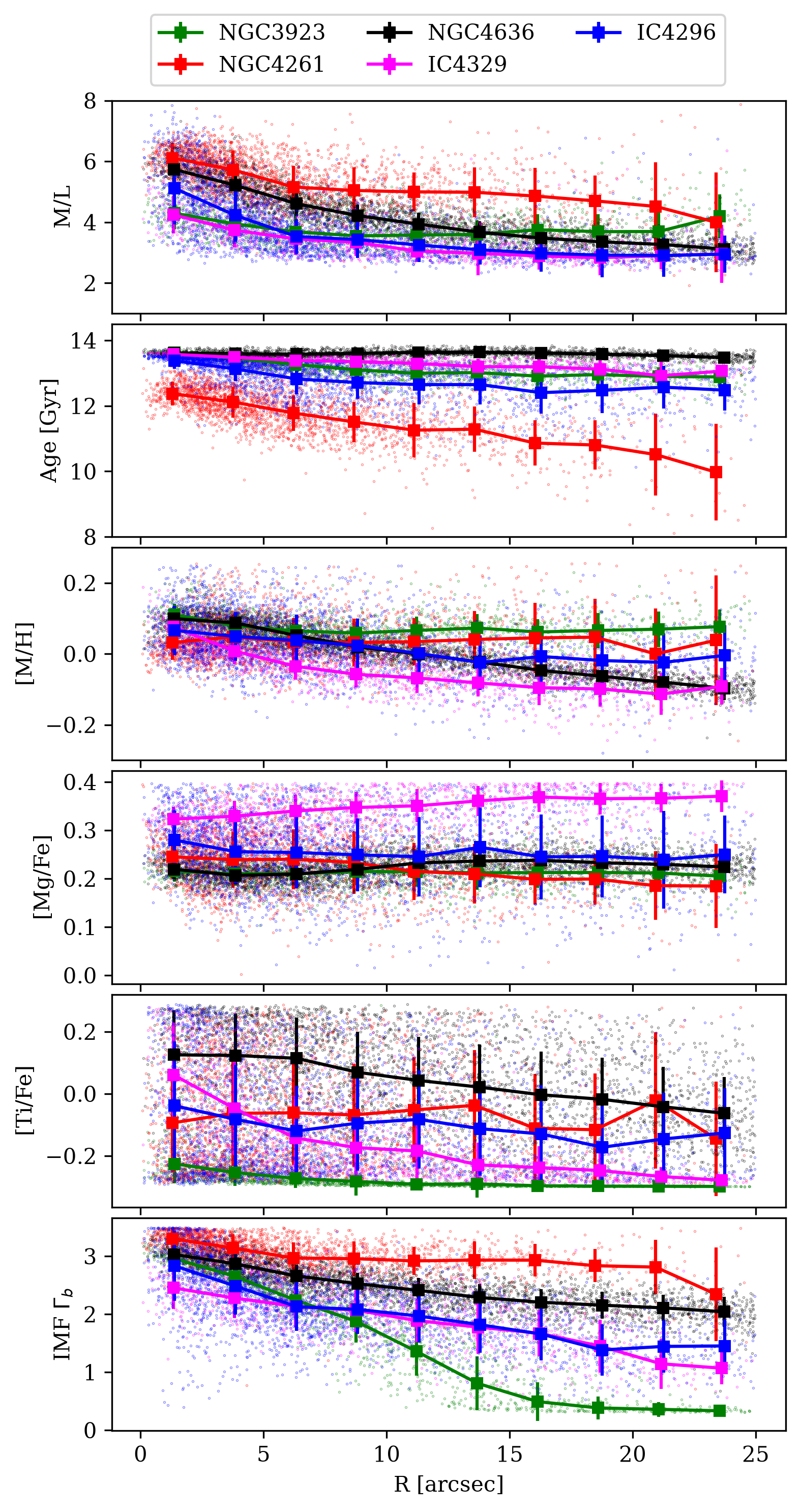}
    \caption{Measured stellar population properties of our sample galaxies from the FIF analysis as a function of radial distance. Squares denote the radially binned data points, while small dots represent individual spatial bins. From top to bottom, the panels display Age, metallicity ([M/H]), [Mg/Fe], [Ti/Fe], and the IMF slope ($\rm \Gamma_{b}$).}
    \label{fig:1Dfif_maps}
\end{figure}

\subsection{DYNAMITE model}\label{sec4.2}

Orbit based superposition technique has been quite successful in measuring BHs masses in the central regions of nearby galaxies and is thoroughly explained in the literature. We used the publicly available Schwarzschild triaxial modelling software, DYnamics, Age, and Metallicity Indicators Tracing Evolution \citep[DYNAMITE v4.0;][]{Jethwa2020, Thaternov2022}. The triaxial dynamical modelling process consists of three major steps: a) First, for a given system, a gravitational potential model is constructed. b) Second, Based on the assumed gravitational potential, a set of orbit libraries is derived. c) The third step involves projecting these orbits into the observed space and finding the best combination that matches the observed kinematics and luminosity. In the following, we describe these steps in detail and explain the model's parameters.

The model gravitational potential consists of three components: a central BH, stellar mass, and a dark matter halo. The contribution of the central BH to the total potential is represented as a point source using a Plummer potential. The stellar potential is derived using the MGE described in Section~\ref{Sec4:Stellar_mass_model} and is deprojected into three-dimensional space according to three viewing angles ($\rm \theta$, $\rm \phi$, $\rm \psi$). These angles are intrinsically connected to the triaxial shape of the galaxy ($p$, $q$, and $u$), where $q=C/A$, $p=B/A$, and $u=A'/A$, with $A$, $B$, and $C$ representing the intrinsic major, intermediate, and minor axes, respectively, and $A'$ the projected major axis. In our modelling, the individual MGE components are scaled by the radially varying $\rm M/L$ values derived from the stellar mass-to-light ratio maps. The stellar potential is therefore generated from the same underlying MGE light distribution, with the mass normalization of the components modified according to the corresponding $\rm M/L$ values.

For the dark matter halo, we used the Navarro–Frenk–White (NFW; \cite{Navarro1996}) radial profile. Numerical cosmological simulations predict that the DM halo can be well represented by a cuspy DM profile, which is accurately described by the NFW halo. In our gravitational potential model, the NFW profile is parametrized by two parameters: $\rm f_{DM}$ = $\rm M_{200}/M_{*}$, the fraction of dark matter within $\rm R_{200}$ (the virial radius, at which the density is 200 times the critical density of the Universe), and the concentration parameter $\rm c_{200}$, defined as $\rm c_{200} = \rm R_{200}/R_s$, where $\rm R_s$ is the scale radius of the NFW profile. We used \cite{Dutton2014} concentration mass relation to reduce the NFW parametrization to one parameter. In summary, our gravitational potential model has six parameters in total : $\rm M_{BH}, S, p, q, u \ and \ f_{DM}$.

For each underlying gravitational potential model, one orbit library is generated, which consists of different types of orbits (see below). These orbits are uniquely characterized by three integrals of motion: binding energy (E), angular momentum (I2), and a third non-classical conserved integral (I3). We broadly sample the orbits in two categories: First, tube orbits, which include both short- and long-axis tubes, sampled by E, I2, and I3; and second, box orbits, which are sampled by E and two spherical angles ($\rm \theta$ and $\rm \phi$). The number of tube orbit libraries sampled with combinations of (E, I2, and I3) was nE, nI2, nI3 = 35 $\times$ 15 $\times$ 11, while box orbits were sampled similarly with E, $\rm \theta$ and $\rm \phi$. To produce a smooth orbital distribution, each orbit was dithered by a factor of 3 by perturbing the initial conditions. Finally, a non-negative least-squares (NNLS) fit is used to assign weights to the individual orbits such that their combination, after accounting for the PSF and aperture binning, reproduces the observed line-of-sight velocity distribution ($V$, $\sigma$, $\rm h_3$, and $\rm h_4$) derived from the SINFONI and large-scale MUSE observations. The intrinsic 3D and projected 2D aperture masses derived from the stellar-mass MGE models are included as constraints in the NNLS fit to ensure that the stellar distribution is reproduced to within 2\% \citep{Vandenbosch2008}.

Once we have explored the parameter space, we select minimum $\rm \chi^{2}$ models as the best fit model from the $\rm \chi^{2}$ distribution. In the Schwarzschild modelling, $\rm \chi^{2}$ fluctuates significantly with the standard deviation of $\sim$ $\sqrt{2 \times \rm N_{kin}}$ \citep{Vandenbosch009, Lipka2021}, where $\rm N_{kin}$ is the number of observables (spatial bins of kinematics maps) making it difficult to apply the classical statistical 1 $\rm \sigma$ confidence interval, which is determined by $\rm \Delta \rm \chi^{2}$ = 1 for one degree of freedom. To estimate the uncertainties in our model parameters, we follow the approach used in \cite{Vandenbosch2008, Vandenbosch009, Zhu2018}. Specifically, we define the 1 $\sigma$ confidence interval as $\Delta \chi^{2}$ = $\rm (\chi^{2}- \rm \chi^{2}_{min})/\sqrt{2 \times \rm n_{GH}\times \rm N_{kin}}$, where $\rm n_{GH}$ is the number of Gauss-hermite parameters and $\rm N_{kin}$ is number of spatial bins. 

\input{All_parameters_appendix}

\section {Results} \label{Sec5:Results}

In this section, we first present our results for the stellar populations of the galaxies, followed by the dynamical mass modelling results.

\subsection{Stellar population properties}

Figure \ref{fig:2Dfif_maps} presents the 2D stellar population maps for each galaxy in our sample, obtained from the FIF analysis. From top to bottom, each row shows the luminosity weighted stellar age, metallicity [M/H], [Mg/Fe], [Ti/Fe] and IMF slope ($\rm \Gamma_{b}$), respectively, for each galaxy as labelled on the top panel. All five galaxies in our sample show an old stellar population, a typical behaviour observed for the massive galaxies. NGC3923, NGC4636, IC4296, and IC4329 host predominantly old stellar populations, with ages of approximately 12–14 Gyr. NGC4261 shows a somewhat younger stellar population and a radial age gradient, decreasing from approximately 12 Gyr in the central region to about 10 Gyr at larger radii. Looking at the spatial maps (Figure \ref{fig:2Dfif_maps}) we note that for NGC3923, NGC4261 and IC4296 the outer bins extending beyond 20 arcsec have unusually different values which are attributed to low SN, in these bins which also correspond to all other measured quantities. 

In the metallicity [M/H] maps (second row in Figure \ref{fig:2Dfif_maps}), NGC3923 shows a slight metallicity gradient in the central region, peaking at 0.1 dex, and otherwise remains fairly constant out to 25 arcsec. NGC4261 maintains a fairly constant metallicity within 20 arcsec, followed by a steep decline in one half of the map, while the other half shows an increase in metallicity. This behaviour is likely due to the lower signal-to-noise ratio (S/N) and the impact of telluric absorption in the spectrum. An inspection of the spectral fits in these bins reveals strong skylines, indicating that the metallicity values beyond 20 arcsec are likely unreliable. NGC4636 exhibits a metallicity gradient, with the metallicity peaking at 0.1 dex in the centre and decreasing to -0.1 dex in the outskirts. IC4296 shows a slight metallicity gradient in the central region, peaking at 0.1 dex, and otherwise remains fairly constant out to 25 arcsec. IC4329 also shows a steep metallicity gradient, with [M/H] peaking at around 0.1 dex in the central region and decreasing towards larger radii.
 
For all galaxies we find that [Mg/Fe], which is used as a proxy for the star formation timescale, remains constant within 25 arcsec. Compared to previous studies on the [Mg/Fe] parameters of massive galaxies \citep{Navarro2018}, we find no correlation between age and [Mg/Fe]. Only in NGC4636 we find a slight, almost negligible anti-correlation in the inner 10 arcsec between age and [Mg/Fe] parameter. 

We notice that for our sample galaxies, the IMF constraints are primarily driven by the gravity-sensitive TiO absorption features. As illustrated in Figure~\ref{fig:exm_fifline}, the Fe and Mg features are reproduced similarly by models with different IMF slopes, whereas the largest differences between the free-IMF, Kroupa-like, and Salpeter models occur in the TiO bands. For NGC3923 and IC4329, the TiO2 feature is strongly affected by residual skyline contamination and was therefore excluded from the FIF analysis. For these two galaxies, the IMF constraints are consequently based mainly on TiO1 together with the remaining fitted features. We notice that all galaxies in our sample, except NGC4261 and IC4296, show a slight gradient in [Ti/Fe], whereas NGC4261 and IC4296 show relatively flat profiles.

Regarding the IMF maps, all five galaxies in our sample display a bottom-heavy IMF in their central regions, with the IMF slope generally decreasing towards larger radii. This behaviour is consistent with previous findings for massive ETGs. NGC4261 shows the most extreme behaviour, maintaining a bottom-heavy IMF both in the central and outer regions. NGC3923 displays a bottom-heavy IMF within the central 10 arcsec, followed by a steep radial decline to sub-Kroupa values in the outer regions. This sharp decrease may be influenced by the exclusion of the TiO2 feature from the FIF analysis and should therefore be interpreted with caution, as it may reflect the reduced IMF sensitivity of the fit rather than the intrinsic IMF behaviour of NGC3923. In contrast, NGC4636, IC4296, and IC4329 show declining IMF slopes, transitioning towards Milky Way-like or Kroupa-like values in their outer regions.

Figure \ref{fig:1Dfif_maps} shows the 1D radially binned stellar population quantities, depicting the comparison between the stellar population properties of each galaxy. In Figure \ref{fig:1Dfif_maps}, we included spatial bins out to 25 arcseconds to ensure a minimum S/N ratio of 50 per bin. As observed in the 2D spatial maps, we clearly see a flat stellar age for all galaxies except NGC4261, where we observe a radial gradient in age, ranging from $\sim$12 to $\sim$10 Gyr. The lower panel in Figure \ref{fig:1Dfif_maps} shows the 1D radial IMF profiles, which clearly exhibit radial variations in the IMF.

%### FIG 6
\begin{figure}
\centering
        \includegraphics[width=1.0\linewidth]{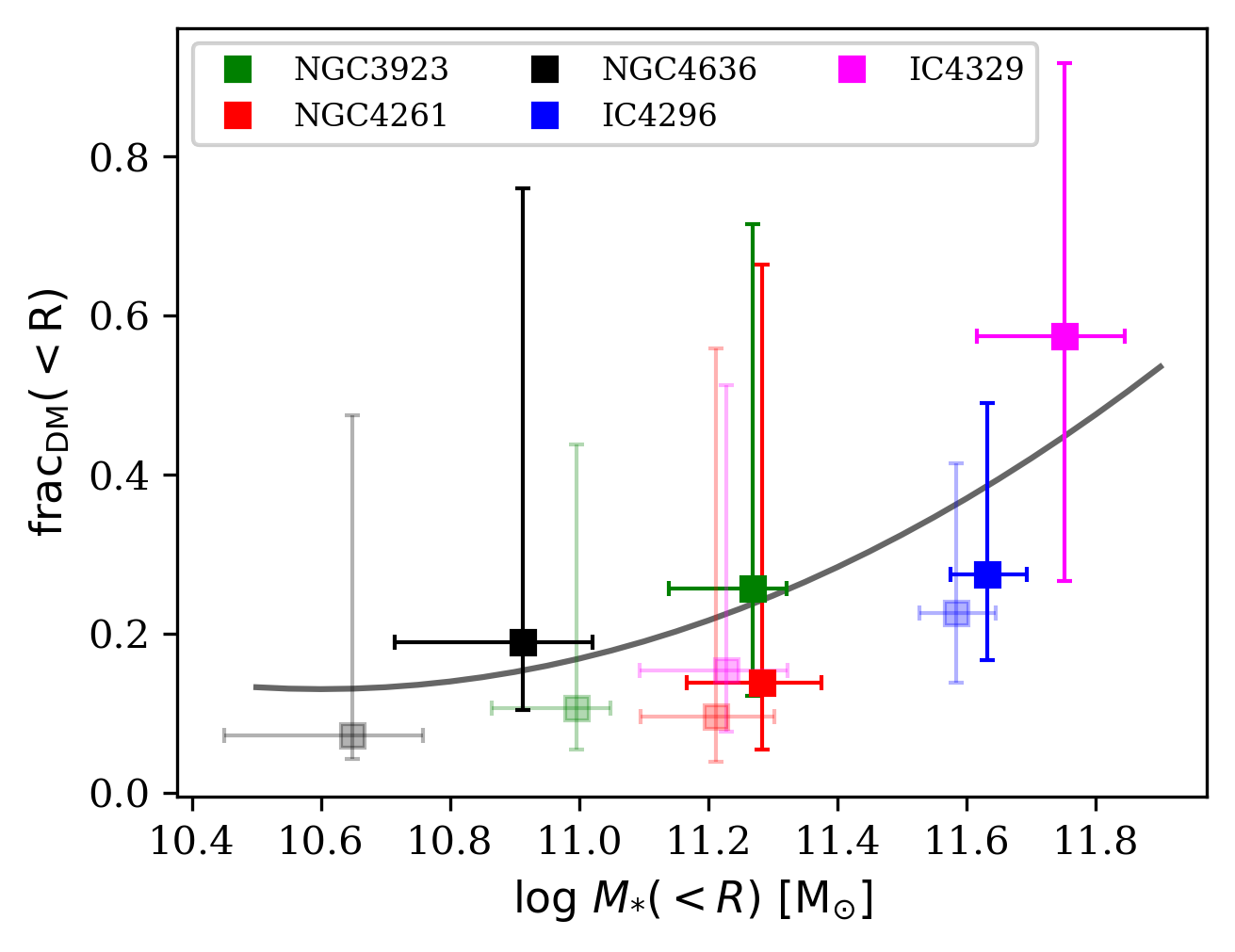}
    \caption{Fraction of dark matter, $\rm frac_{DM}=M_{DM}/M_{total}$, as a function of the enclosed stellar mass for our sample galaxies. Filled squares show the DM fraction and stellar mass measured within 1 $\rm R_{eff}$, while the faint squares show the corresponding values within the maximum radial extent of the MUSE data. The error bars indicate the full range spanned by the models within the adopted $3\sigma$ confidence region. The grey line shows the relation between $\rm frac_{DM}$ and stellar mass within 1 $\rm R_{eff}$ from \citet{Cappellari2013DM}.}
    \label{fig:Results_DMfraction}
\end{figure}

\subsection{DYNAMITE setup and modelling results} \label{Sec5.2:Results}

We construct the Schwarzschild models as described in Section \ref{sec4.2}. Our modelling setup includes a total of six parameters: the intrinsic shape parameters p, q, and u; the dark matter (DM) halo fraction $\rm f_{DM}$; the stellar mass scaling factor S, defined as the ratio of the dynamical stellar $\rm M_{dyn}/L$ to the stellar $\rm M_{*}/L$ ratio; and the central black hole (BH) mass, $\rm M_{BH}$. The stellar $\rm M_{*}/L$ ratio is known, derived from the FIF analysis, and incorporated into the stellar mass models (Section \ref{Sec4:Stellar_mass_model}), whereas $\rm M_{dyn}/L$ is not known and is fitted as part of the model. The parameter $\rm f_{DM}$ represents the DM contribution to the gravitational potential, while $\rm M_{dyn}/L$ corresponds purely to the dynamical mass associated with the stellar component, excluding the DM contribution.
Since the parameter S is the ratio between the dynamical ($\rm M_{dyn}/L$) and stellar ($\rm M_{*}/L$) mass-to-light ratios, it is expected to be close to one. Any significant deviation from this value indicates uncertainties either in the stellar population modelling of $\rm M_{*}/L$ or in the DM parametrization. It is important to note that the parameter S is not exactly the same as the mismatch parameter or the mass-excess factor ($\rm \alpha$) for the IMF used in studying IMF variation in the dynamical modelling \citep{Cappellari2012, Smith2020}. The mass-excess factor is defined as the ratio of the mass-to-light ratio obtained from dynamical methods to the reference Milky Way (Kroupa-like) mass-to-light ratio derived from appropriate stellar population models. In our case, the reference stellar $\rm M_{*}/L$ used in S is obtained from the FIF analysis considering a variable IMF, rather than a fixed IMF as in the case of the $\alpha$ parameter.

We modelled the MUSE and SINFONI data in two steps. First, we modelled the MUSE and SINFONI kinematics together for each galaxy, keeping all parameters free, to constrain the intrinsic shape and gravitational potential of the galaxy. In the second step, we utilized the best-fit parameter values—namely, $p$, $q$, $u$, and $\rm f_{DM}$—determined in the first step to perform combined modelling of the SINFONI and MUSE kinematics. During this step, we varied only the BH mass and $S$ over a finer grid to derive a precise estimate of the BH mass. In the first step of the modelling, the starting parameter values were $p = 0.7$, $q = 0.4$, $u = 0.8$, $\log \rm f_{DM} = 4.0$, and $S = 1$. For the BH mass, we used a starting value expected from the $\rm M_{BH}-\sigma$ relation \citep{VandenBosch2016}. For each galaxy, we ran more than 1000 models until the minimum $\rm \chi^{2}$ was reached. Table \ref{Tab1:results} shows the best-fit values of all the parameters obtained from the first-step modelling of the MUSE and SINFONI kinematics. Note that Table \ref{Tab1:results} also shows the black hole mass values, which are not constrained in this step and therefore do not represent the final best-fit values. Only for NGC3923 we measure the BH mass from the MUSE data, and this value is very close to that obtained from our combined SINFONI and MUSE modelling. As discussed in Thater2026, the MUSE data for NGC3923 are of very good quality and resolve the sphere of influence of its black hole. Therefore, the BH mass of NGC3923 derived from the MUSE data is very close to that obtained from the combined SINFONI and MUSE data (see Section \ref{Sec5:ResultsBHmass}). The Figure \ref{fig:Appendix_chi2_map} in the Appendix shows the full-grid $\rm \chi^{2}$ distribution obtained from the first step of the modelling of the MUSE and SINFONI kinematics for each galaxy.

We found that all the galaxies in our sample exhibit tri-axial geometry as expected for the massive galaxies. The detailed results of the intrinsic shape and orbital structure of the modelled galaxies will be presented in a subsequent paper (Chaturvedi et al., in prep). Here, we primarily focus on the derived BH masses of the galaxies. 

%### FIG 7
\begin{figure*}
\centering
        \includegraphics[width=1.00\linewidth]{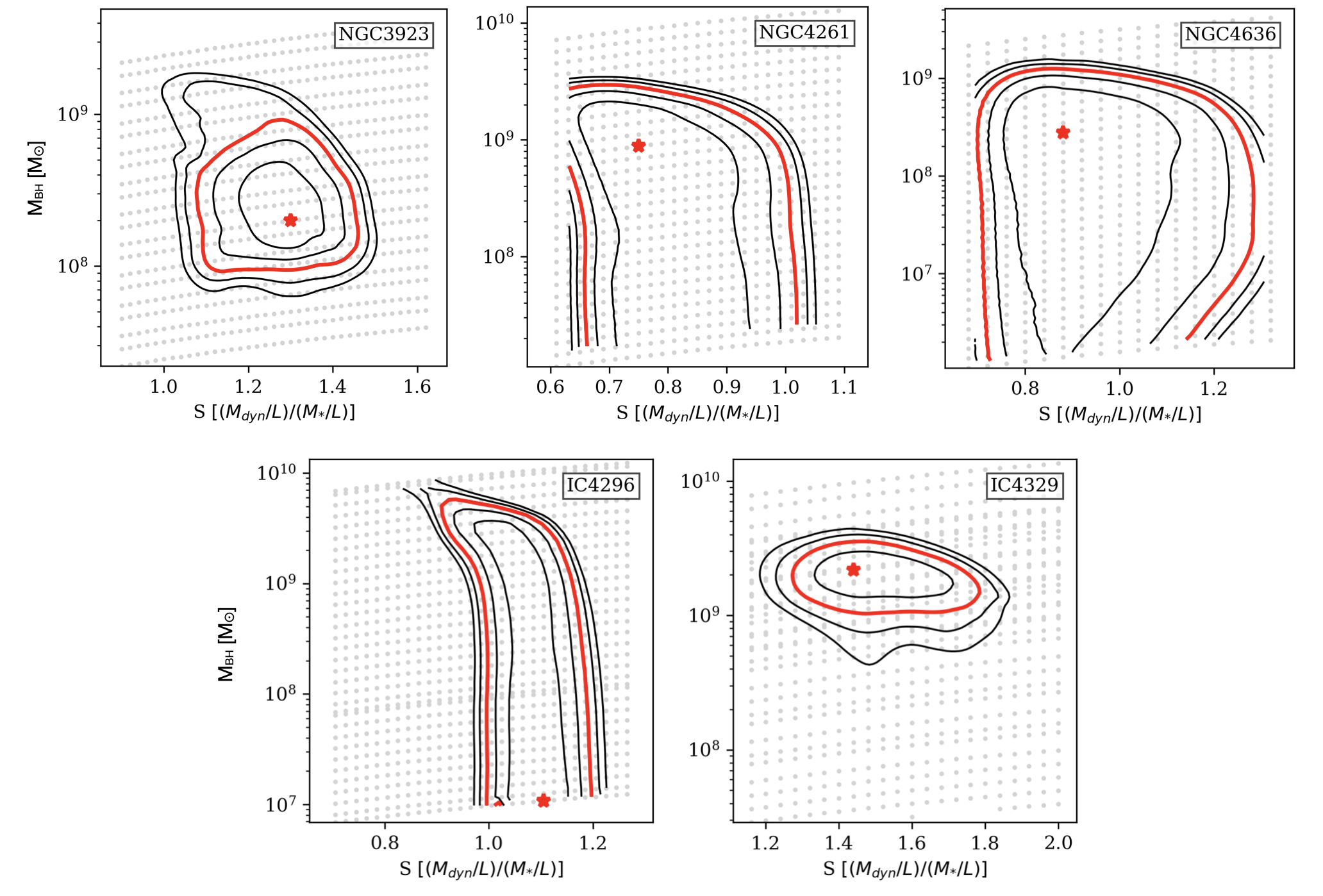}
    \caption{Results of the triaxial Schwarzschild modelling conducted with DYNAMITE for the joint fitting of MUSE and SINFONI kinematics. Each grey dot corresponds to an individual model for different S and $\rm M_{BH}$ values, while the other parameters remain fixed (see text for details). The best-fitting model, derived as the minimum of the $\chi^{2}$ distribution is marked with the red star. Overlaid contours denote the $\rm \Delta \chi^{2} = (\chi^{2} -\chi^{2}_{min})/\sqrt{2 \times \rm n_{GH}\times \rm N_{kin}}$ levels,  where $\rm n_{GH}$ is the number of Gauss-Hermite parameters and $\rm N_{kin}$ is the number of spatial bins. Red contours indicate 3$\sigma$ confidence interval.}
    \label{fig:Results_svsmbh}
\end{figure*}

\subsubsection{Enclosed mass profile and fraction of DM}

In Appendix \ref{fig:Appendix_maps}, Figures \ref{fig:AppendixkinmapNGC3923}, \ref{fig:AppendixkinmapNGC4261}, \ref{fig:AppendixkinmapNGC4636}, \ref{fig:AppendixkinmapIC4296}, and \ref{fig:AppendixkinmapIC4329} show the comparison between the observed and modelled MUSE and SINFONI kinematic maps of NGC3923, NGC4261, NGC4636, IC4296, and IC4329, respectively. The comparison indicates that our models generally reproduce the observed kinematics well. Examining the residuals, the discrepancy between the data and the models is typically at the level of 1--2$\rm \sigma$, with some bins at the outer edges reaching around 3$\rm \sigma$. However, for NGC4636, our models do not fully reproduce the $\rm h_{3}$ measurements in an annular region between $\sim$5--10'', coinciding with the region of strong kinematic twists and outer long-axis rotation. This could be related to systematic uncertainties in the extraction of $\rm h_{3}$, or possibly to the complex LOSVD in the central region, for which a Gauss-Hermite parametrisation might not be optimal. However, the residuals in the other kinematic moments are generally smaller.

We use the best-fit DM halo parameters to derive the enclosed mass profiles of the gal
axies, displayed in Figure \ref{fig:massprofile}. The vertical solid and dashed grey lines in Figure \ref{fig:massprofile} indicate the maximum radial extent of the MUSE observations and one effective radius ($\rm R_{eff}$), respectively. Although the maximum radial extent of the MUSE observations for all galaxies in our sample lies within 1 $\rm R_{eff}$, the enclosed mass profiles enable us to estimate the dark-matter fraction ($\rm frac_{DM}=M_{dark}/M_{tot}$) within 1 $\rm R_{eff}$ for these galaxies. Figure \ref{fig:Results_DMfraction} shows the DM fraction within the maximum radial extent of the MUSE data (faint squares) as a function of the enclosed stellar mass $\rm M_{*}$ for all galaxies. We additionally show the corresponding values within 1 $\rm R_{eff}$ with filled squares. The green curve in Figure \ref{fig:Results_DMfraction} represents the relation between $\rm frac_{DM}$ and stellar mass within 1 $\rm R_{eff}$, assuming an NFW DM halo, as reported by \cite{Cappellari2013DM}. We adopt the $\rm R_{eff}$ values for our sample galaxies from the literature, mainly from \cite{Cappellari2013DM} and \cite{Saglia2016}.

The best-fitting DM fractions within the maximum radial extent of the MUSE data are relatively modest for all five galaxies, ranging from approximately 7\% to 23\%. When extrapolated to 1 $\rm R_{eff}$, we measure $\rm frac_{DM}$ values of approximately 26\%, 14\%, 19\%, 27\%, and 57\% for NGC3923, NGC4261, NGC4636, IC4296, and IC4329, respectively. The large range of DM fractions allowed by the models within our adopted 3$\rm \sigma$ confidence region indicates that the DM contribution is only weakly constrained.

For NGC3923, previous dynamical modelling based on globular cluster kinematics by \citet{Norris2012} reported a DM fraction of approximately 17.5\% within 1 $\rm R_{eff}$. However, their adopted effective radius of 53.4 arcsec in the B band differs from the value of 86.1 arcsec adopted in our analysis. At our adopted 1 $\rm R_{eff}$, we measure a best-fitting DM fraction of approximately 26\%. For a more direct comparison, at 53.4 arcsec we obtain $\rm frac_{DM}=0.36^{+0.43}_{-0.19}$. Although the best-fitting value is higher than the measurement of \citet{Norris2012}, their measured value lies within the range allowed by our models at the $3\sigma$ confidence level.

Overall, the best-fitting DM fractions of our galaxies are broadly consistent with previous estimates. However, the large ranges of DM fractions allowed by the models within our adopted $3\sigma$ confidence region indicate that the DM halo remains relatively weakly constrained, while the enclosed stellar mass is considerably better constrained.

\input{BHmass}

%### Figure 8 ####
\begin{figure*}
\centering
        \includegraphics[width=0.95\linewidth]{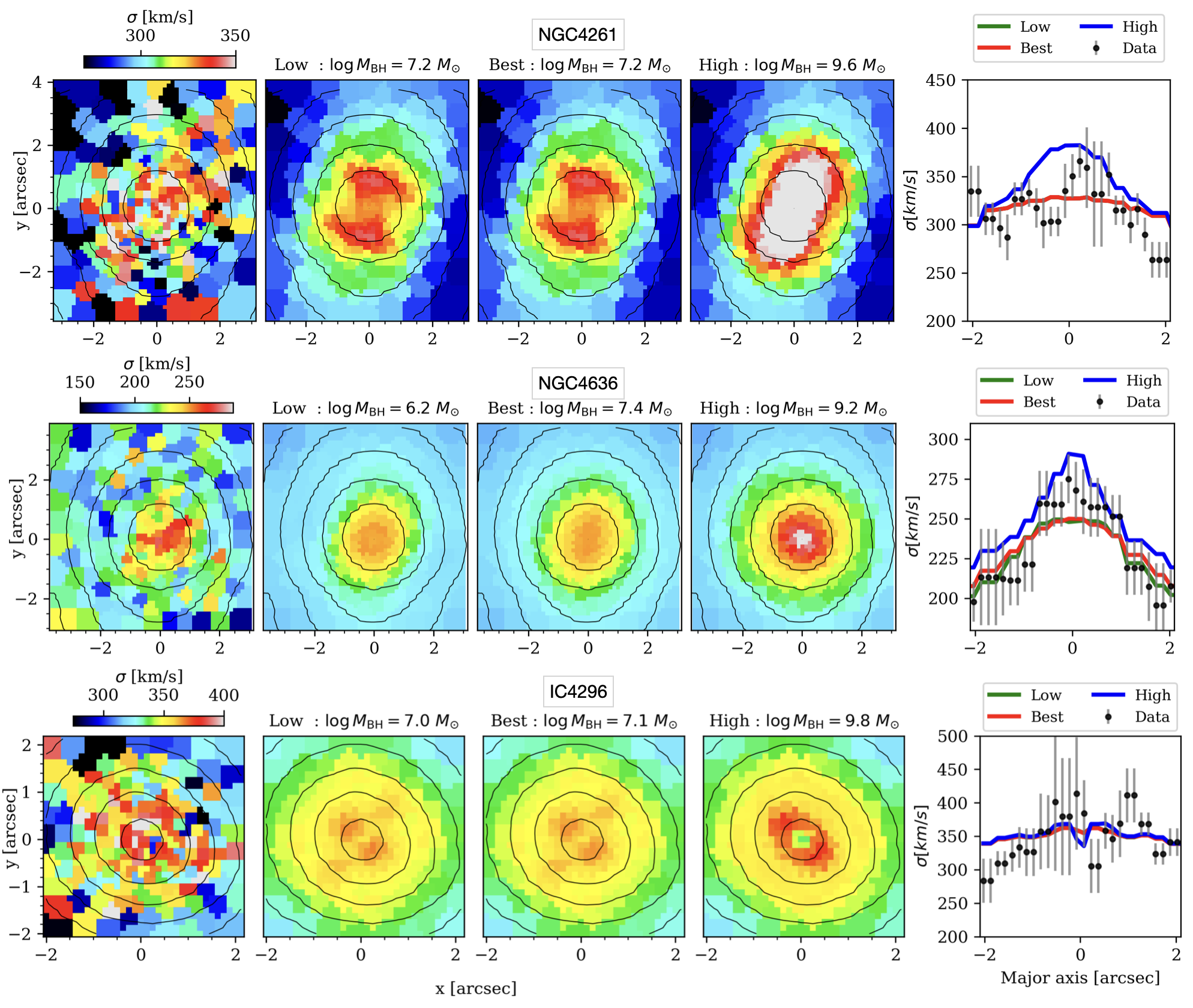}
    \caption{Comparison of the velocity dispersion maps from the SINFONI data with those from the Schwarzschild models for different BH masses for NGC4261 (top row), NGC4636 (middle row), and IC4296 (bottom row). In each row, from left to right, the panels show the observed velocity dispersion map and the model velocity dispersion maps corresponding to the low, best-fit, and high BH masses. For each galaxy, these models are shown at the best-fit $\rm S$ parameter as listed in Table~\ref{Tab2:results}. The corresponding BH masses are indicated at the top of each subpanel.} The last column shows a comparison of the observed and modelled velocity dispersion profiles along the major axis.
    \label{fig:Results_ngc4261_ngc4636_mbh}
\end{figure*}

\subsubsection{Black hole Mass}\label{Sec5:ResultsBHmass}

After estimating the intrinsic shape and DM halo parameters, we rerun DYNAMITE to model the SINFONI and MUSE kinematics simultaneously over a refined grid of $\rm M_{BH}$ and S. Figure \ref{fig:Results_svsmbh} illustrates the $\Delta \chi^{2} = (\chi^{2} - \chi^{2}_{\text{min}})/\sqrt{2 \times \rm n_{GH}\times \rm N_{kin}}$ distribution between $\rm M_{BH}$ and S parameters for the galaxies. Each dot in Figure \ref{fig:Results_svsmbh} represents a model and red star indicates the best-fit estimate, while the red contour marks the 3$\rm \sigma$ confidence interval. The contours in Figure \ref{fig:Results_svsmbh} are smooth by a Gaussian kernel with a width of 1$\sigma$. For our BH mass measurements, we report uncertainties at the 3$\sigma$ confidence level. 

For NGC3923, we determine a BH mass of $1.99^{+6.76}_{-1.05} \times 10^{8} \ \rm M_{\odot}$ and a scaling factor of $\rm S = 1.30^{+0.15}_{-0.21}$, indicating that the dynamical model prefers a slightly higher stellar $\rm M_{*}/L$ normalization than predicted by our stellar population models. For NGC4261, our modelling yields an upper limit on the BH mass of $28.27 \times 10^{8} \ \rm M_{\odot}$, with $\rm S = 0.75^{+0.23}_{-0.23}$. Similarly, for NGC4636, we obtain an upper limit on the BH mass of $12.40 \times 10^{8} \ \rm M_{\odot}$, with $\rm S = 0.88^{+0.37}_{-0.18}$. The measured scaling factors for NGC4261 and NGC4636 are close to unity, suggesting that the stellar mass normalizations inferred from our stellar population and dynamical modelling analyses are broadly consistent. For IC4296, our modelling also yields an upper limit on the BH mass of $33.57 \times 10^{8} \ \rm M_{\odot}$, with $\rm S = 1.11^{+0.10}_{-0.10}$. In the case of IC4329, we determine a BH mass of $21.78^{+13.58}_{-11.32} \times 10^{8} \ \rm M_{\odot}$ and a scaling factor of $\rm S = 1.44^{+0.30}_{-0.16}$. The scaling factor for IC4329 is the highest among the five galaxies, indicating that the dynamical model prefers a larger stellar $\rm M_{*}/L$ normalization than predicted by the stellar population models. Table \ref{Tab2:results} summarizes the best-fit values of $\rm M_{BH}$ and $\rm S$ for our sample galaxies.

We compare our BH mass estimates with previous measurements available in the literature and with the results obtained by \citet{Thater2026}. Previously, \citet{Saglia2016} utilized SINFONI data for NGC3923 and performed axisymmetric Schwarzschild modelling. For NGC3923, they reported $\rm M_{BH} = 2.81 \pm 0.097 \times 10^{9} \ M_{\odot}$ and $\rm M/L \sim 4.22$ in the z-band. Our measured $\rm M_{BH}$ is approximately a factor of 14 lower than that estimated by \citet{Saglia2016}. The confidence intervals of the two measurements do not overlap, with our upper bound remaining approximately a factor of 3 lower than the lower bound reported by \citet{Saglia2016}. For IC4296, our modelling yields an upper limit on the BH mass of $\leq 3.36 \times 10^{9} \ \rm M_{\odot}$, which is approximately a factor of 2.5 higher than the original measurement of $\rm M_{BH} = 1.34^{+0.21}_{-0.19} \times 10^{9} \ M_{\odot}$ reported by \citet{Bonta2009}. Their BH mass measurement was obtained using dynamical modelling of the nuclear ionized-gas kinematics observed with HST/STIS. \citet{Saglia2016} later rescaled this measurement to their adopted distance for IC4296, obtaining $\rm M_{BH} \simeq 1.30 \times 10^{9} \ M_{\odot}$. The previous BH mass measurement of IC4296 lies within the upper limit derived from our modelling. These differences between our results and the previous measurements may arise from several factors. For NGC3923, the previous measurement was obtained using axisymmetric orbit-based modelling, whereas our analysis employs triaxial Schwarzschild modelling. For IC4296, the previous measurement of \citet{Bonta2009} was instead based on ionized-gas dynamical modelling. Our analysis uses stellar kinematics and includes a DM halo and a spatially varying stellar $\rm M_{*}/L$, further distinguishing our modelling approach from the previous measurements.

For NGC3923, \citet{Thater2026} reported $\rm M_{BH} = 11.9^{+13.4}_{-9.0} \times 10^{8} \ M_{\odot}$, which is approximately a factor of 6.1 higher than our measured value. For NGC4261, our $3\sigma$ upper limit is approximately a factor of 1.27 higher than the corresponding upper bound from \citet{Thater2026}, who reported $\rm M_{BH} = 11.4^{+10.8}_{-9.5} \times 10^{8} \ M_{\odot}$. Similarly, for NGC4636, our $3\sigma$ upper limit is approximately a factor of 1.6 higher than the corresponding upper bound from \citet{Thater2026}, who reported $\rm M_{BH} = 4.7^{+3.0}_{-4.3} \times 10^{8} \ M_{\odot}$. In contrast, for IC4296, our upper limit is lower by approximately a factor of 2.0 than the corresponding upper bound from the \citet{Thater2026} measurement of $\rm M_{BH} = 35.1^{+33.7}_{-25.7} \times 10^{8} \ M_{\odot}$. Finally, for IC4329, our measured BH mass is approximately a factor of 1.14 lower than the value reported by \citet{Thater2026}, $\rm M_{BH} = 24.3^{+15.3}_{-16.5} \times 10^{8} \ M_{\odot}$. We discuss the implications of these differences, particularly the effects of adopting a constant or spatially varying $\rm M_{*}/L$, further in Section~\ref{Sec6:discussion_imfbh}.

To investigate how well our models reproduce the inner SINFONI kinematics, in Figure \ref{fig:Results_ngc4261_ngc4636_mbh}, we present a comparison between the SINFONI velocity dispersion map and those generated using the low mass BH, current best-fit BH mass and the high mass BH for NGC4261, NGC4636, and IC4296, derived from their $\chi^2$ distribution maps. We observe that for NGC4261 (first row in Figure \ref{fig:Results_ngc4261_ngc4636_mbh}), the best-fit BH mass can reproduce the observed kinematic profile, whereas the high BH mass shows a clear peak in the central region. The SINFONI kinematics for NGC4261 reveal strong asymmetry in the kinematic peak along the major axis. This could potentially be related to the PSF or low-quality S/N data. Given the low quality of the SINFONI data, with velocity dispersion errors of approximately $\sim$10 \%, it is challenging to further investigate the kinematics or place tighter constraints on the BH mass. Notably, the upper limit on the BH mass we derive for NGC4261 is consistent with the value reported by \cite{Ruffa2023}, where they used CO dynamical modelling and estimated a BH mass of $\rm 1.62 \pm 0.04 \times 10^{9} \ M_{\odot}$ at the 1$\rm \sigma$ confidence level. However, for our Schwarzschild models, the SINFONI data are not of high enough quality to account for the peak observed in the velocity dispersion profile, as would be expected for the \cite{Ruffa2023} BH mass. 

For NGC4636 (second row in Figure \ref{fig:Results_ngc4261_ngc4636_mbh}), both the best-fit and high BH masses effectively reproduce the observed kinematics of NGC4636, showing no significant differences. As with NGC4261, the SINFONI kinematics for NGC4636 have large uncertainties in the velocity dispersion. While a central peak is observed, the velocity dispersion profile is relatively flat and exhibits abrupt bin-to-bin variations, making it inconsistent with the smoother peak expected from a high mass BH. The third row in Figure \ref{fig:Results_ngc4261_ngc4636_mbh} shows the velocity dispersion maps for IC4296, where we find that both the low-mass and best-fit BH models reproduce the SINFONI velocity dispersion maps equally well. In comparison with NGC 4636 and NGC 4261, IC 4296 shows a flatter velocity dispersion profile, with large uncertainties in the central bins, therefore, the models show little difference and reproduce the data equally well. In conclusion, the SINFONI data for NGC4261, NGC4636 and IC4296 are not inconsistent with a high-mass BH, but they also lack the quality required to drive the models toward high mass BH. Furthermore, as discussed in Thater2026, NGC4261 and NGC4636 show complex stellar kinematic structures. Our triaxial models reproduce the observed kinematics reasonably well, with most of the normalized residuals remaining within $\pm3$ times the observational uncertainties. However, different combinations of orbital families can reproduce these complex kinematic structures over a range of BH masses, reducing the sensitivity of the models to $\rm M_{BH}$. Given the combination of complex stellar kinematics and the limited quality of the data, our dynamical models are therefore unable to robustly determine a best-fitting BH mass for these galaxies, and we instead report upper limits.

%### Figure 9 ###
\begin{figure*}[h]
\centering
        \includegraphics[width=1.0\linewidth]{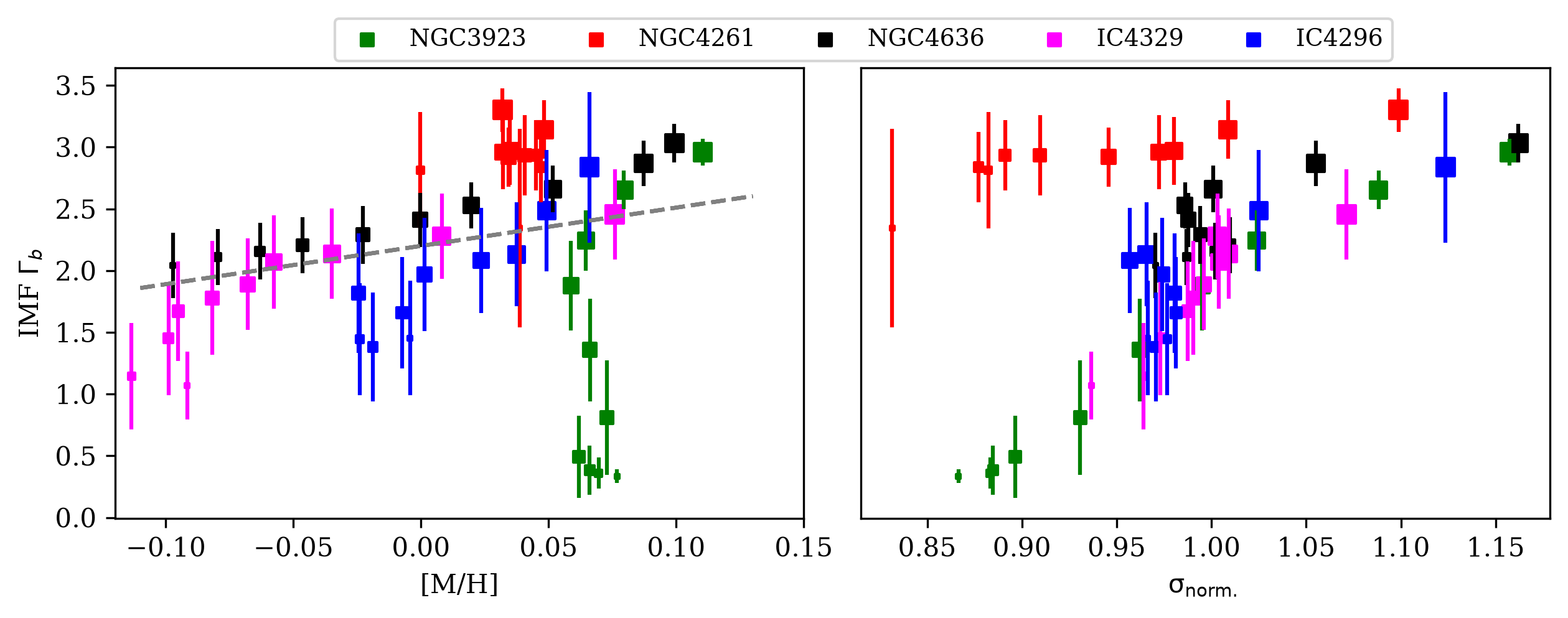}
    \caption{Relation between IMF Slope ($\rm \Gamma_{b}$), metallicity [M/H], and velocity dispersion ($\rm \sigma $) of Galaxies: Left Panel: The slope of the IMF is shown as a function of metallicity, derived from the FIF analysis. The dashed line represents the IMF–[M/H] relation from \cite{Navarrojun2015}. Larger squares indicate bins closer to the centers of the galaxies. Right Panel: The slope of the IMF is displayed as a function of normalized velocity dispersion.}
    \label{fig:Discussion_imf}
\end{figure*}

\section{Discussion} \label{Sec6:Discussion}

\subsection{Physical drivers behind IMF variation}

The physical drivers behind the spatial variation of the IMF in ETGs are still not fully understood. Several studies have reported correlations between IMF variation and galaxy properties such as velocity dispersion, [Mg/Fe], and metallicity. Considering the global properties of massive galaxies, \citet{Conroy2012} reported that the IMF mass-normalization factor correlates more strongly with [Mg/Fe] than with velocity dispersion. In contrast, \citet{Cappellari2013} found that the IMF mass normalization, determined using dynamical modelling for 30 galaxies in common with the sample of \citet{Conroy2012}, correlates more strongly with velocity dispersion. Even after accounting for aperture effects in deriving these quantities, the different approaches do not necessarily lead to the same correlations \citep{McDermind2014}. These results indicate that the physical parameter most closely connected to IMF variation is still under debate.

Metallicity has also been suggested as an important parameter regulating the IMF. Using the CALIFA sample, \citet{Navarrojun2015} found a strong correlation between IMF slope and metallicity, which was further supported by \citet{VanDokkum2017}. However, subsequent spatially resolved studies have shown that the IMF-metallicity relation is more complex. Based on spatially resolved IFU observations of FCC167, \citet{Navarro2019} found that the central regions follow the IMF-metallicity relation proposed by \citet{Navarrojun2015}, while the outer regions show a considerably larger scatter. Similarly, \citet{Navarro2021} found for galaxies in the Fornax cluster that the scatter in the IMF-metallicity relation increases towards lower metallicities.

Following these studies, we investigate the relation between IMF slope, metallicity, and velocity dispersion for our sample galaxies. Figure~\ref{fig:Discussion_imf} presents the IMF slope ($\rm \Gamma_{b}$) as a function of [M/H] (left panel) and normalized velocity dispersion, $\rm \sigma_{norm}$ (right panel), using radial bins within 25 arcsec for which the spectra have S/N higher than 50. The velocity dispersion is normalized by the median velocity dispersion of each galaxy. The dashed line in the left panel of Figure~\ref{fig:Discussion_imf} represents the IMF-metallicity relation from \citet{Navarrojun2015}. The size of the squares decreases with increasing radial distance from the galaxy centre.

The relation between IMF slope and metallicity varies significantly among the galaxies. NGC4636, IC4296, and IC4329 show a strong positive relation between IMF slope and metallicity, with Spearman correlation coefficients of 1.00, 0.855, and 0.927, respectively. Their central regions also lie relatively close to the IMF--metallicity relation of \citet{Navarrojun2015}, while a larger scatter is observed towards lower metallicities and larger radii. In contrast, NGC3923 and NGC4261 show weak or no correlation between IMF slope and metallicity, with Spearman coefficients of 0.176 and 0.018, respectively. For NGC3923, although the IMF shows a strong radial gradient, the metallicity becomes relatively constant after its initial decline, resulting in a weak overall IMF-metallicity relation. A similar behaviour is observed for NGC4261, where the IMF slope varies radially while [M/H] remains relatively constant at approximately 0.05 dex. NGC4261 also differs from the other galaxies in having a somewhat younger stellar population, which becomes progressively younger towards larger radii.

Besides metallicity, velocity dispersion is another galaxy property that has frequently been associated with IMF variation. The velocity dispersion measured globally, for example within $\sim1\,\rm R_{eff}$, or within the central regions has been shown to correlate with the IMF in several studies \citep{Treu2010, Conroy2012, Cappellari2013, LaBarbera2015}. Based on MUSE observations of M87, \citet{Sarzi2018} found that the IMF correlates with velocity dispersion, although they argued that velocity dispersion itself does not necessarily drive the radial IMF variation. Our spatially resolved stellar population measurements allow us to investigate this relation within each of our sample galaxies. We measure the radially binned velocity dispersion in the same radial bins used for the stellar population parameters shown in Figure~\ref{fig:1Dfif_maps}. The right panel of Figure~\ref{fig:Discussion_imf} shows the IMF slope as a function of velocity dispersion normalized by the median value of each galaxy.

The IMF-$\sigma$ relation also varies from galaxy to galaxy. NGC3923 and NGC4261, which show weak or no IMF--metallicity correlation, instead show strong positive correlations between IMF slope and velocity dispersion, with Spearman correlation coefficients of 1.00 and 0.952, respectively. IC4329 also shows a strong IMF-$\sigma$ correlation, with a Spearman coefficient of 0.964. At the same time, IC4329 shows a similarly strong correlation between IMF slope and metallicity ($\rho=0.927$), indicating that for this galaxy the IMF slope is closely associated with both quantities. In contrast, NGC4636 and IC4296 show weaker IMF-$\sigma$ correlations, with Spearman coefficients of 0.60 and 0.285, respectively, while both galaxies show much stronger correlations between IMF slope and metallicity.

Overall, our results show that no single galaxy property exhibits a uniformly strong correlation with the IMF across the entire sample. For NGC3923 and NGC4261, the IMF is more strongly correlated with velocity dispersion than with metallicity, whereas for NGC4636 and IC4296 the IMF is more strongly correlated with metallicity. IC4329 shows a strong correlation with both quantities. We also find that the central regions generally follow more ordered IMF-metallicity trends, while the scatter becomes larger towards the outer regions. This behaviour is qualitatively consistent with the spatially resolved results of \citet{Navarro2019} and \citet{Navarro2021}.

The difference between the central and outer regions may also be connected to the formation histories of massive ETGs. Galaxies in our sample lie at the high-mass end of the galaxy population and are generally expected to undergo a two-phase formation scenario \citep{Oser2010}. Their central regions are thought to form predominantly at high redshift through rapid, dissipative in-situ star formation, whereas their outer regions are subsequently built through the accretion of stars and major or minor merger events. Such different formation pathways can introduce a larger diversity of stellar populations in the outer regions and may contribute to the increased scatter observed in the IMF relations at larger radii. This interpretation is consistent with previous studies \citep{Navarrofeb2015, Sarzi2018, Navarro2021, Navarro2024}, which have also found tighter IMF relations in galaxy centres and larger scatter towards the outer regions. Although our sample consists of only five galaxies, our results suggest that IMF variation is likely connected to a combination of local stellar population properties and galaxy assembly history rather than to a single universal parameter. Extending this analysis to a larger sample of ETGs will be important for further understanding the physical mechanisms driving IMF variation.

%### Figure 10 ###
\begin{figure*}
\centering
        \includegraphics[width=1.0\linewidth]{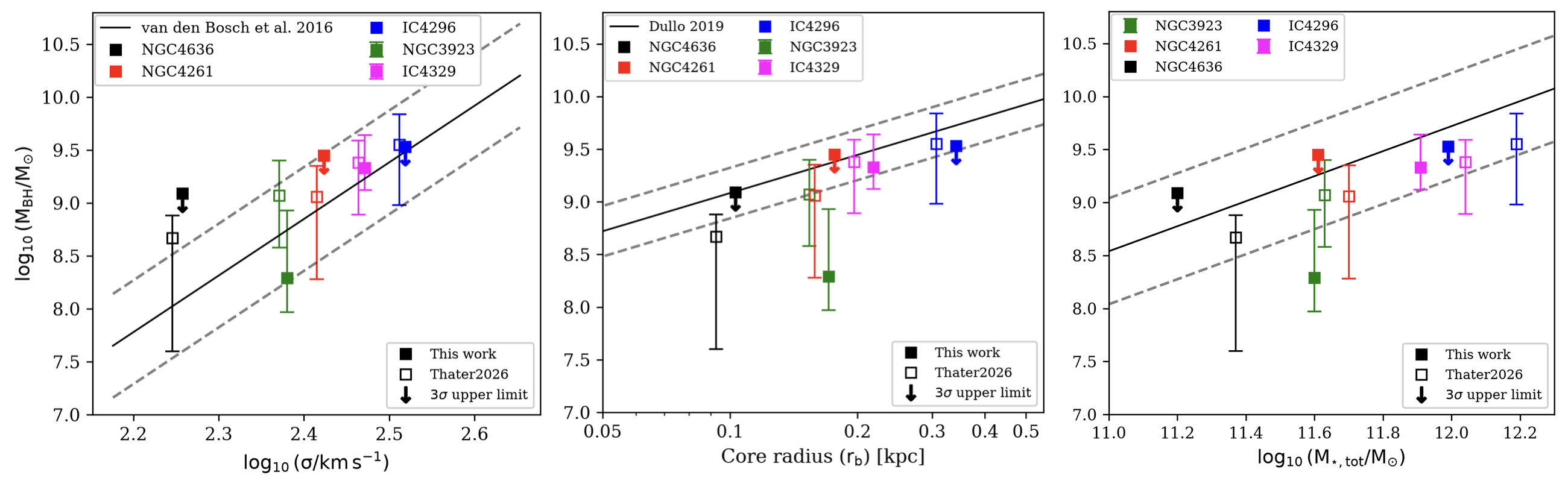}
        \caption{Comparison of the measured BH masses for our five galaxies with the $\rm M_{BH}$--$\rm \sigma$ relation (left panel), $\rm M_{BH}$--$ \rm r_b$ relation (middle panel), and $\rm M_{BH}$--$\rm M_{*,tot}$ relation (right panel). Filled squares show the BH masses derived in this work using a spatially varying stellar $\rm M_{*}/L$, while open squares show the corresponding measurements from Thater2026 obtained assuming a constant $\rm M_{*}/L$. Downward arrows indicate the $3\sigma$ upper limits obtained for NGC4261, NGC4636, and IC4296. For clarity, Thater2026 measurements are shifted slightly along the x-axis: by $-5~{\rm km/s}$ in velocity dispersion in the left panel and by 10\% towards smaller core-radius values in the middle panel. $\rm M_{BH}-\sigma$ relation left panel is from \cite{VandenBosch2016}, $\rm M_{BH}-r_{b}$ relation in the middle panel is from \cite{Dullo2019} and in the right panel the $\rm M_{BH}-M_{*}$ relation is taken from \cite{Shankar2019}.}
    \label{fig:Discussion_Mbh_sigma_rb}
\end{figure*}

\subsection{Scaling relation between BH mass and galaxy properties}
We examined how our BH mass measurements compare with the established scaling relations between BH mass and galaxy properties. We first compared our measurements with the $\rm M_{BH}$--$\sigma$ relation. The left panel of Figure~\ref{fig:Discussion_Mbh_sigma_rb} shows the positions of our sample galaxies on the $\rm M_{BH}$--$\sigma$ relation adopted from \citet{VandenBosch2016}. The dashed lines indicate the $1\sigma$ intrinsic scatter of the relation. Open square symbols indicate the BH masses measured by \citet{Thater2026}. Based on our $\rm M_{BH}$ measurements, we find that the BH masses of NGC3923 and IC4329 are consistent with the $\rm M_{BH}$--$\sigma$ relation within their uncertainties and the intrinsic scatter of the relation. For NGC4261, NGC4636, and IC4296, for which we obtain only upper limits, the BH masses expected from the $\rm M_{BH}$--$\sigma$ relation lie below our upper limits and are therefore allowed by our dynamical constraints.

Next, we compared our BH mass measurements with the predictions from the $\rm M_{BH}$--core-radius ($\rm r_b$) relation. Massive elliptical galaxies are thought to grow through mergers, and mergers involving SMBHs can leave a distinct signature in their central regions in the form of a light-deficit core. Several studies have demonstrated a strong correlation between $\rm M_{BH}$ and $r_b$ \citep{Rusli2013, Dullo2014, Thomas2016, Dullo2019}. The middle panel of Figure~\ref{fig:Discussion_Mbh_sigma_rb} shows the positions of our sample galaxies on the $\rm M_{BH}$--$r_b$ scaling relation adopted from \citet{Dullo2019}. The dashed lines indicate the $1\sigma$ intrinsic scatter of the relation.

We obtained core-radius measurements for our sample galaxies from the literature, except for NGC3923, for which no previous measurement is available. To determine the core radius of NGC3923, we extracted its one-dimensional isophotal surface-brightness profile from the HST-WFC3/F814W image using the Python photutils package. Following the method of \citet{Krajnovic2020}, we fitted the surface-brightness profile with a two-component model consisting of an inner core-S\'{e}rsic profile and an outer exponential halo. Figure~\ref{fig:Appendix_NGC3923_rb} shows the best-fitting model and the corresponding residuals. We obtain a core radius of $r_b = 1.679 \pm 0.23''$ ($0.171 \pm 0.05$ kpc). The measured core size of NGC3923 is comparable to those typically observed in massive early-type galaxies and is consistent with the presence of a depleted stellar core \citep{Dullo2019}. Using the BH mass derived by \citet{Thater2026} under the assumption of a constant stellar $\rm M/L$, NGC3923 lies close to the observed $\rm M_{BH}$--$r_b$ relation. In contrast, our best-fitting BH mass lies below the $1\sigma$ intrinsic scatter of the relation. For IC4329, our measured BH mass is consistent with the relation within its uncertainty. For NGC4261 and NGC4636, the BH masses expected from the relation lie below our upper limits and are therefore allowed by our constraints. The upper limit for IC4296 lies below the mean relation but remains within its intrinsic scatter. We also find that the BH masses measured by \citet{Thater2026} are broadly consistent with the $\rm M_{BH}$--$\rm r_b$ relation, with NGC4636 lying below the $1\sigma$ intrinsic scatter but remaining consistent within its measurement uncertainty.

In the right panel of Figure~\ref{fig:Discussion_Mbh_sigma_rb}, we compare our BH mass measurements with the $\rm M_{BH}$--$\rm M_{\star,\mathrm{tot}}$ relation, using the total stellar masses derived from our variable-$\rm M_{*}/L$ models. We adopt the observed $\rm M_{BH}$--$M_{\star,\mathrm{tot}}$ relation for local early-type quiescent galaxies presented by \citet{Shankar2019}. The dashed lines indicate a scatter of 0.5 dex around the relation. NGC3923 shows the largest offset from the mean relation, with its best-fitting BH mass lying below the 0.5 dex scatter, although its upper uncertainty extends into the scatter of the relation. The BH mass measured for IC4329 is consistent with the relation within its uncertainty. For NGC4261 and NGC4636, the BH masses expected from the scaling relation lie below our upper limits and are therefore allowed by our dynamical constraints. The upper limit for IC4296 lies below the mean relation but remains within the adopted scatter. For comparison, the BH masses measured by Thater2026, plotted using their corresponding total stellar masses, are broadly consistent with the $\rm M_{BH}$--$\rm M_{\star,\mathrm{tot}}$ relation within their measurement uncertainties and the adopted scatter.

\subsection{Impact of variable IMF on BH mass} \label{Sec6:discussion_imfbh}

Finally, we compare our BH mass measurements with those derived by Thater2026 assuming a constant stellar mass-to-light ratio. Our analysis uses the same kinematic datasets and the same underlying MGE representation of the stellar light distribution as Thater2026, with the main difference being the inclusion of a radially varying $\rm M_{*}/L$ derived from the stellar population analysis, thereby accounting for spatial variations in the IMF and other stellar population properties. As seen in the left panel of Figure~\ref{fig:Discussion_Mbh_sigma_rb}, the effect of adopting a variable $\rm M_{*}/L$ is not uniform across the sample. For NGC3923 and IC4329, for which we obtain constrained BH masses, the inferred BH masses are approximately 83\% and 11\% lower, respectively, than those obtained assuming a constant $\rm M_{*}/L$. For NGC4261 and NGC4636, our BH mass upper limits are approximately 27\% and 60\% higher than the corresponding upper bounds from Thater2026, whereas for IC4296 our upper limit is approximately 51\% lower. These differences illustrate that a spatially varying stellar $\rm M_{*}/L$ can affect both the inferred BH mass and the BH mass constraints, without producing a systematic change in the same direction across the sample.

In our dynamical modelling, the contributions from the stellar component, dark matter halo, and central BH are modelled simultaneously. The stellar $\rm M_{*}/L$ therefore plays an important role in determining how the mass in the central regions is distributed between the stellar component and the BH. By incorporating a variable $\rm M_{*}/L$, derived from stellar population modelling with a free IMF, we obtain an additional constraint on the radial stellar mass distribution. All galaxies in our sample exhibit bottom-heavy IMFs in their central regions, corresponding to an enhanced contribution from low-mass stars and hence higher stellar $\rm M_{*}/L$ values towards the centre. For example, if the variable-$\rm M_{*}/L$ model assigns relatively more stellar mass to the central region, part of the mass that would otherwise be attributed to the BH can instead be accounted for by the stellar component, resulting in a lower inferred BH mass. However, the effect on the inferred BH mass depends on how the resulting stellar mass profile differs from that assumed in the corresponding constant-$\rm M_{*}/L$ model.

The magnitude and direction of this effect, however, differ substantially among the galaxies. The strongest difference in the measured BH masses is found for NGC3923, where the inferred BH mass decreases by a factor of about six, or approximately 83\%. The considerably smaller decrease of approximately 11\% for IC4329 demonstrates that the magnitude of this effect can vary substantially between galaxies. For NGC4261 and NGC4636, where we obtain only upper limits, the allowed BH mass range instead extends to higher values than the corresponding upper bounds from \citet{Thater2026}, whereas the upper limit for IC4296 becomes considerably tighter. This indicates that changes in the stellar mass profile can also alter the range of BH masses permitted by the dynamical models rather than simply shifting the preferred BH mass to lower values. Previous studies have also demonstrated that accounting for stellar population-driven $\rm M_{*}/L$ gradients can substantially affect dynamical BH mass measurements. For example, \citet{Thater2023} implemented a spatially varying $\rm M_{*}/L$ in the nucleated ETG FCC~47 and found an approximately 50\% decrease in the inferred $\rm M_{BH}$. Similar effects have been investigated in other dynamical modelling studies \citep{McConnel2016, Thater2019}.

Our findings suggest that the impact of adopting a variable $\rm M_{*}/L$ may be more complex and dependent on specific galaxy characteristics and modelling choices. These results highlight that the effect of a variable $\rm M_{*}/L$ on $\rm M_{BH}$ is not necessarily uniform across different systems, but rather may vary with the internal structure and stellar population gradients of the galaxies and generally be connected to the formation history of the galaxy. This highlights the importance of applying variable $\rm M_{*}/L$ treatments across a broader range of systems to better understand their influence on BH mass measurements.

\section{Summary and Conclusions} \label{Sec7:Summary}

In this work, we investigated the impact of a spatially varying initial mass function (IMF) on dynamical black hole mass estimates for a sample of five massive early-type galaxies-NGC3923, NGC4261, NGC4636, IC4296, and IC4329—using triaxial Schwarzschild orbit-based modelling of MUSE and SINFONI observations.

We performed stellar population modelling of the MUSE data using the full index-fitting method (Section \ref{Sec3:FIF_model}) to derive the spatially resolved stellar population properties of the galaxies. Our analysis revealed radial variations in the IMF, resulting in corresponding variations in the stellar mass-to-light ratio. We incorporated these spatially varying stellar $\rm M_{*}/L$ profiles, driven by the free IMF into our stellar mass models (Section \ref{Sec4:Stellar_mass_model}). We then carried out the orbit-based dynamical modelling in two main steps. First, we used the large-scale MUSE and SINFONI kinematics together to constrain the intrinsic shape and dark matter halo fraction. In the second step, we simultaneously fitted the SINFONI and MUSE kinematics to constrain the BH masses, keeping the galaxy shape and dark matter halo parameter fixed (Section \ref{Sec5.2:Results}). Our main findings are summarized below.

\begin{enumerate}
    \item From our stellar population analysis, we found that galaxies in our sample host predominantly old stellar populations (Figures \ref{fig:2Dfif_maps} and \ref{fig:1Dfif_maps}), with ages ranging from 12 to 14 Gyr. In contrast, NGC 4261 has a slightly younger stellar population, with an age of approximately 10–12 Gyr. All galaxies in our sample exhibit radial variations in metallicity and an almost constant, enhanced [Mg/Fe] ratio, reaching at least 2–3 times the solar value.

    \item All five galaxies exhibit bottom-heavy IMFs in their central regions, exhibiting a radial gradient with the IMF slope decreasing towards larger radii (Figures \ref{fig:2Dfif_maps} and \ref{fig:1Dfif_maps}). NGC4261 shows the most extreme behaviour, maintaining a bottom-heavy IMF even at larger radii. NGC4636, IC4296, and IC4329 show declining IMF slopes towards Milky Way- or Kroupa-like values in their outer regions. NGC3923 also shows a strong radial decline, reaching sub-Kroupa values at larger radii, although this behaviour should be interpreted with caution because the TiO2 feature was excluded from its FIF analysis owing to skyline contamination.

    \item The relation between the IMF slope and other galaxy properties varies from galaxy to galaxy (Figure~\ref{fig:Discussion_imf}). NGC4636, IC4296, and IC4329 show a clear positive trend between IMF slope and metallicity, whereas this trend is weak or absent in NGC3923 and NGC4261. In contrast, the IMF slope shows a stronger dependence on velocity dispersion in NGC3923, NGC4261, and IC4329, while this dependence is weaker for NGC4636 and IC4296. We also note that, within the central regions, the IMF generally follows the metallicity trends more closely, while a larger scatter is observed in the outskirts. These results indicate that no single parameter shows a uniformly strong correlation with the IMF across the entire sample.

    \item Using triaxial Schwarzschild dynamical modelling with the DYNAMITE code, we place upper limits on the BH masses of NGC4261, NGC4636, and IC4296 of $28.27 \times 10^{8} \ \rm M_{\odot}$, $12.40 \times 10^{8} \ \rm M_{\odot}$, and $33.57 \times 10^{8} \ \rm M_{\odot}$, respectively. For NGC3923 and IC4329, we derive BH masses of $1.99^{+6.76}_{-1.05} \times 10^{8} \ \rm M_{\odot}$ and $21.78^{+13.58}_{-11.32} \times 10^{8} \ \rm M_{\odot}$, respectively (Figure \ref{fig:Results_svsmbh}).  The derived BH masses and upper limits are broadly consistent with the expected range from the considered BH scaling relations (Figure \ref{fig:Discussion_Mbh_sigma_rb}).
    
    \item Compared with the constant-$\rm M_{*}/L$ models of \citet{Thater2026}, incorporating a spatially varying stellar $\rm M_{*}/L$ has a strongly galaxy-dependent impact on the inferred BH masses and upper-limit constraints. For NGC3923 and IC4329, for which we obtain constrained BH masses, the inferred BH masses decrease by approximately 83\% and 11\%, respectively. For NGC4261 and NGC4636, our BH mass upper limits are approximately 27\% and 60\% higher than the corresponding upper bounds from \citet{Thater2026}, whereas for IC4296 our upper limit is approximately 51\% lower. These differences demonstrate that radial variations in the stellar $\rm M_{*}/L$ can have a non-negligible impact on dynamical BH mass measurements, but do not necessarily shift the inferred BH mass or constraint in the same direction for all galaxies.

\end{enumerate}

With this work, we demonstrate the presence of strong radial IMF gradients in a sample of five massive early-type galaxies, with systematically bottom-heavy IMFs in their central regions. Incorporating the corresponding spatially varying stellar $\rm M_{*}/L$ profiles into the dynamical models can substantially change the inferred BH masses and upper-limit constraints relative to models assuming a constant stellar $\rm M_{*}/L$. For the galaxies with constrained BH masses, we find decreases of approximately 83\% and 11\%, while for the galaxies with upper limits the constraints can either become higher or lower. Although the magnitude and direction of this effect vary substantially among individual galaxies, our results demonstrate that stellar population and IMF gradients can represent an important systematic factor in dynamical BH mass measurements. Extending this approach to a larger sample will be necessary to quantify the typical magnitude of this effect and to determine its implications for the observed scaling relations between BH mass and galaxy properties.

\vspace{0.2cm}
\small

\textbf{Acknowledgements:} We would like to thank the anonymous referee for providing important feedback and comments, which have helped to improve the quality of the manuscript. We would also like to thank the Vienna DYNAMITE team, especially Thomas Maindl and Prashin Jethwa, for their help and discussions regarding the use of the DYNAMITE code. The computational work presented in this study was carried out using AIP's NEWTON cluster and the Austrian Scientific Computing (ASC) infrastructure. This research made use of various software including: Astropy (\href{https://www.astropy.org/}{https://www.astropy.org})- a community-developed core Python package for Astronomy \citep{Astropy2013, Astropy2018}, DYNAMITE \citep{Jethwa2020, Thaternov2022}, MGE \citep{Emsellem1994, Cappellari2002}, pPXF \citep{Cappellari2004,Cappellari2017}, vorbin \citep{Cappellari2003}, scipy \citep{Scipy2020}, photutils \citep{phoutilsBradley}, emcee \citep{Emcee2013}.

\bibliographystyle{aa} % style aa.bst
\bibliography{imfvsbh}

\begin{appendix}

\section{Stellar population analysis} \label{fif_analysis}
\begin{figure*}
\centering
    \includegraphics[width=0.90\linewidth]{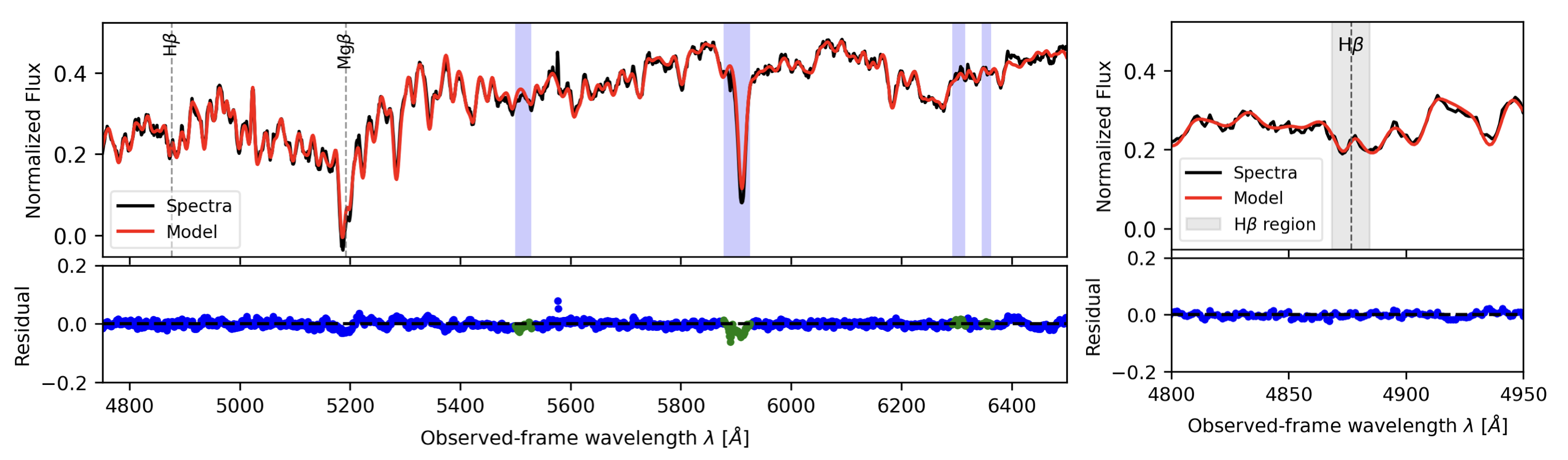}
    \caption{Example pPXF fit from the second step of FIF analysis for the same spatial bin (inner region) of NGC4636. In the left panel red curve shows the best fit spectrum from the pPXF and black curve indicates the observed spectrum. Blue bands show the regions which are masked out from the spectral fitting. The right panel shows the zoomed-in feature around the H$\beta$ line absorption feature.}
    \label{fig:Appendix_fif_step2}
\end{figure*}

Figure~\ref{fig:Appendix_fif_step2} shows an example of the regularised pPXF fit performed in the second step of our stellar population analysis. The fit covers the full wavelength range of 4750--6500\,\AA\ and simultaneously constrains the luminosity-weighted stellar age and metallicity. This wavelength range includes the age-sensitive H$\beta$ absorption feature, which provides an important constraint on the stellar age.

\section{Mass-to-light ratio maps} \label{modelled_map}

\begin{figure*}
\centering
    \includegraphics[width=0.85\linewidth]{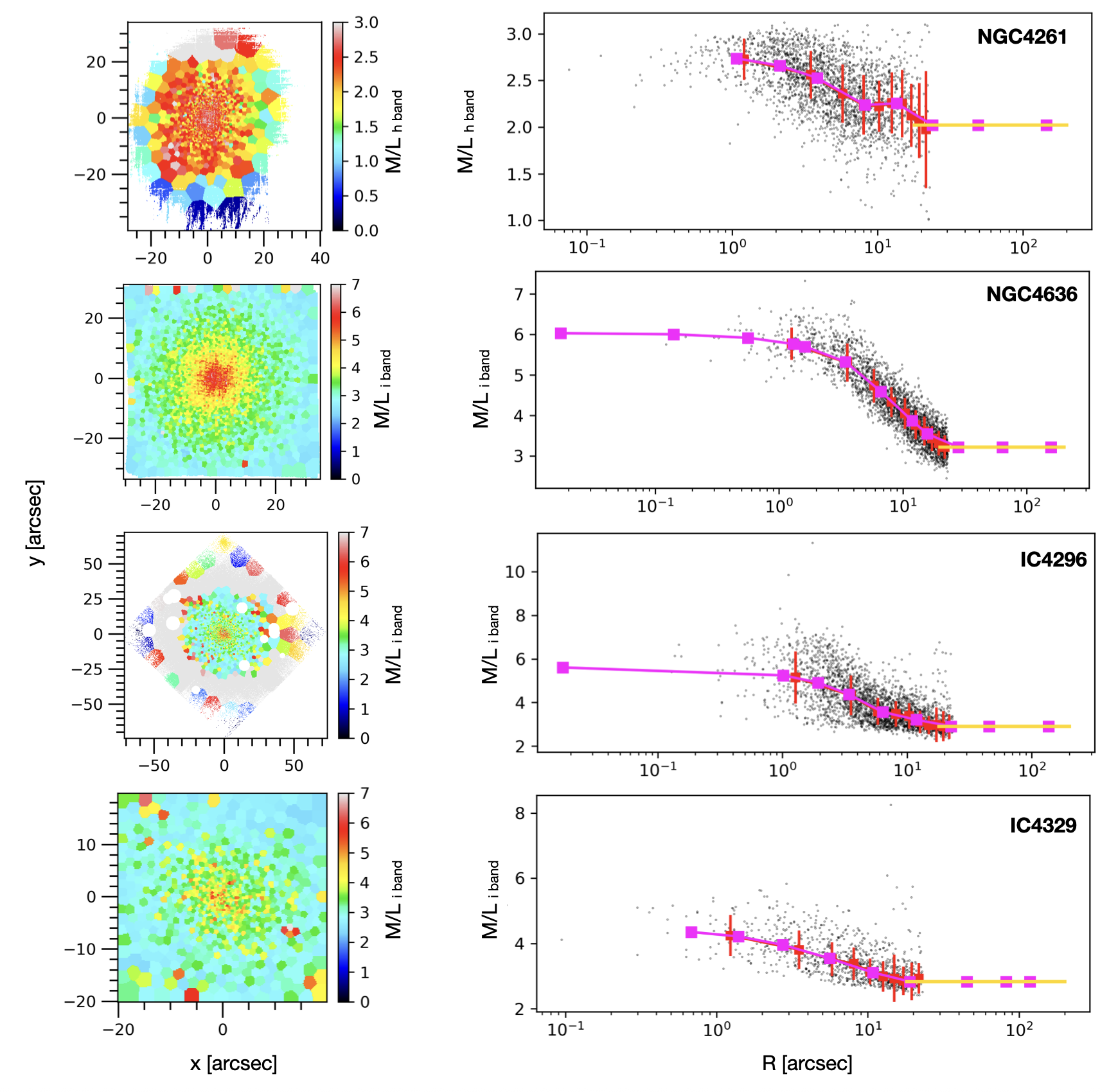}
    \caption{Stellar $\rm M_{}/L$ maps (left panels) and the interpolated and extrapolated radial $\rm M_{}/L$ values (right panels), with extrapolation applied at radii $>20$ arcsec. From top to bottom, the galaxies are NGC4261, NGC4636, IC4296, and IC4329. Red squares denote the radially binned $M_{}/L$ values obtained from the FIF analysis. Grey dots denote the individual spatial bins of the stellar $\rm M_{}/L$ maps. Magenta squares show the interpolated or extrapolated $\rm M_{*}/L$ values evaluated on the same radial grid as the MGE models.}
    \label{fig:Appendix_mge_mlmap}
\end{figure*}

\section{Best-fit DYNAMITE models} 

\begin{figure*}
\centering
        \includegraphics[width=0.90\linewidth]{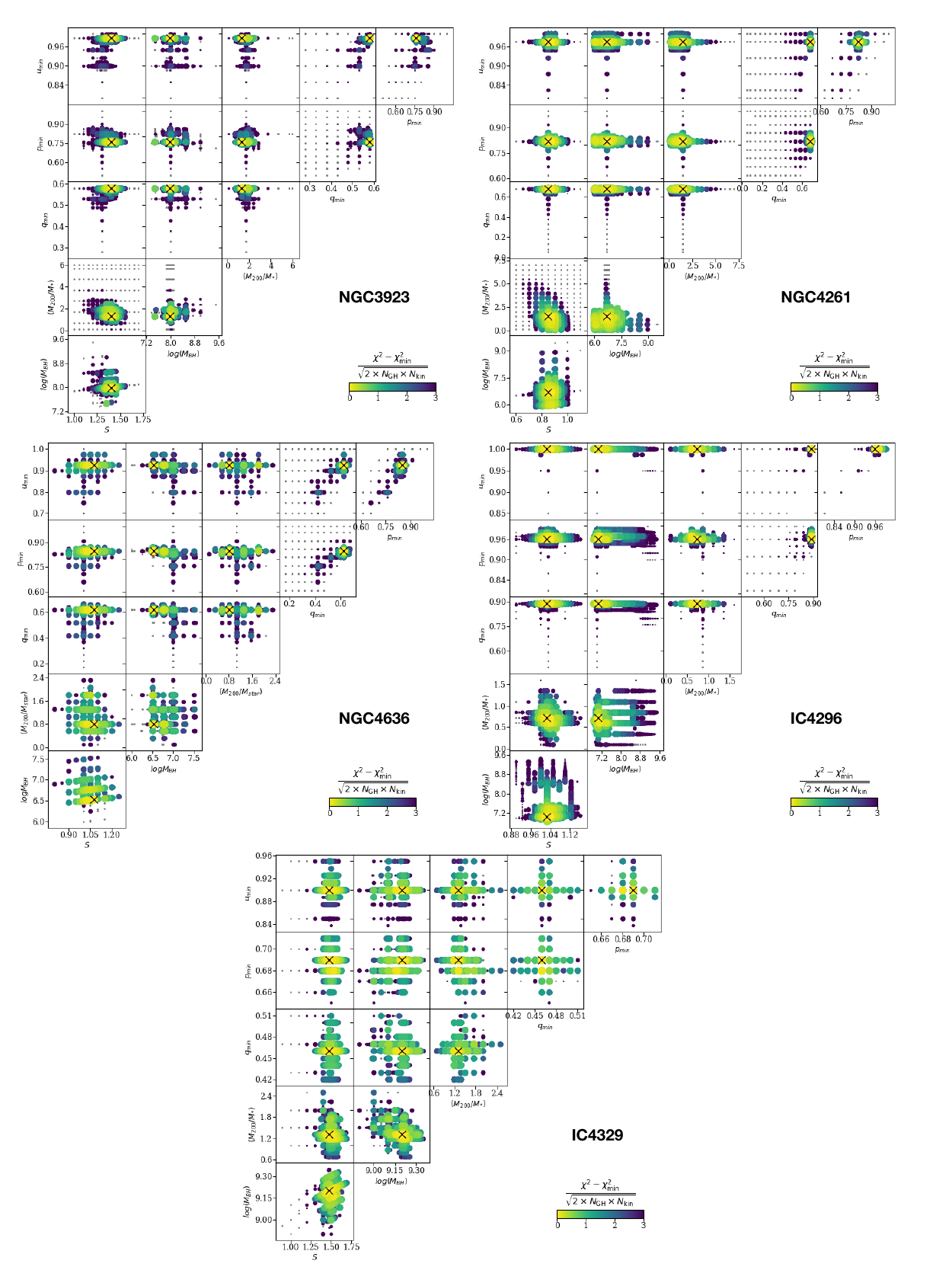}
    \caption{$\chi^{2}$ distribution of our dynamical models obtained from fitting MUSE and SINFONI data together to constrain intrinsic shape and dark matter parameters for each galaxy. The six parameters are S: scaling factor between $\rm M_{dyn}/L$ to $\rm M_{*}/L$; $\rm logM_{BH}$: the black hole mass in logarithmic solar mass units; $f_{DM}$ = $\rm M_{200}/M_{star}$; q: intrinsic minor-to-major axis ratio; p: intrinsic medium-to-major axis ratio and u: the ratio between projected and intrinsic major axis. Each point represents one model, color-coded with the $\Delta \chi^{2}$ = $(\rm \chi^{2}- \rm \chi^{2}_{min})/\sqrt{2 \times \rm n_{GH}\times \rm N_{kin}}$, where $\rm n_{GH}$ is the number of Gauss-hermite parameters and $\rm N_{kin}$ is the number of spatial bins.  A value of $\Delta \chi^{2}$ $<$ 1 indicates the model within 1$\rm \sigma$ confidence interval. The best fit model is denoted by the black cross.  }
    \label{fig:Appendix_chi2_map}
\end{figure*}

\section{Observed and modelled kinematics maps}\label{fig:Appendix_maps}

\begin{figure*}[h]
    \centering
    \subfigure{
        \includegraphics[width=0.51\textwidth]{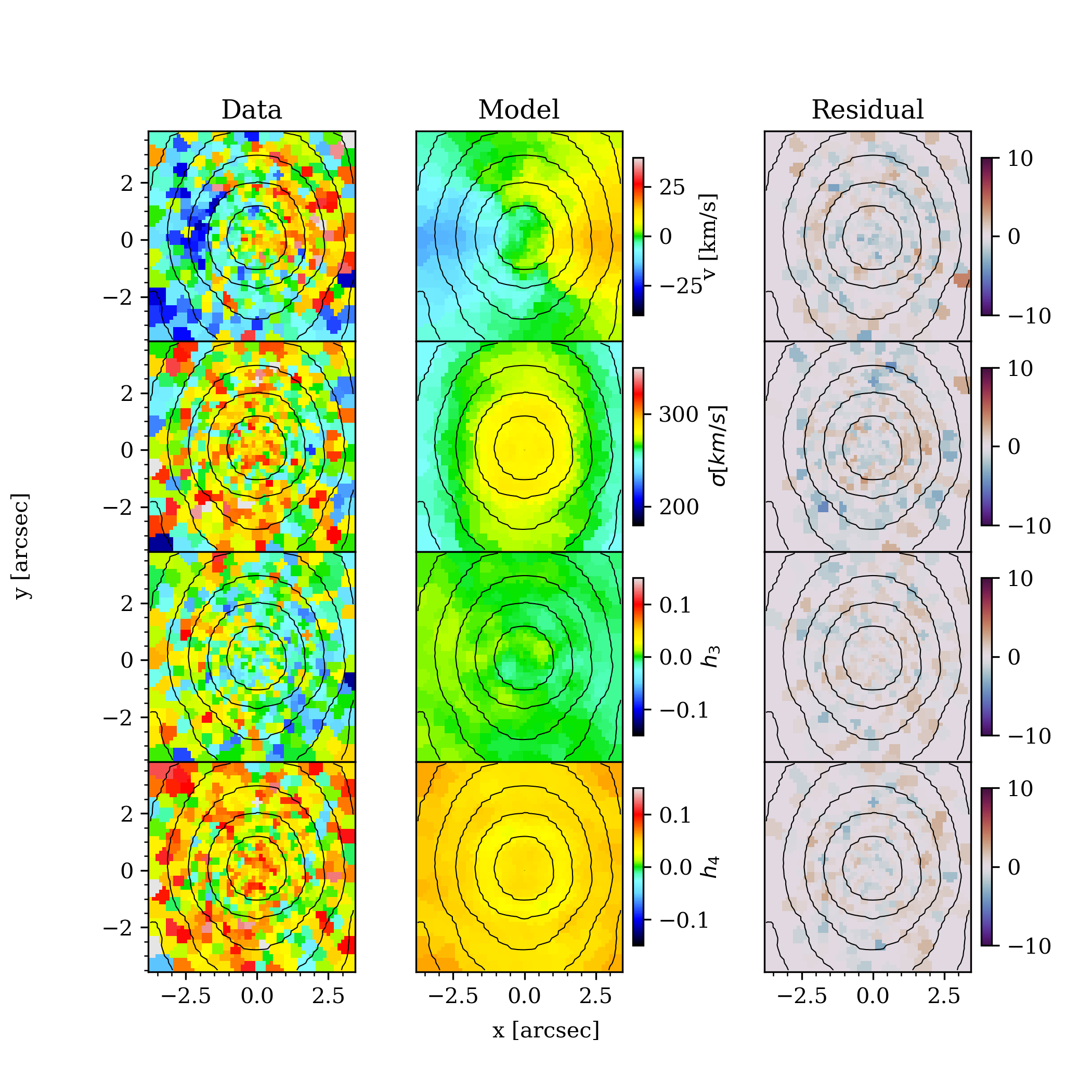}
        \includegraphics[width=0.50\textwidth]{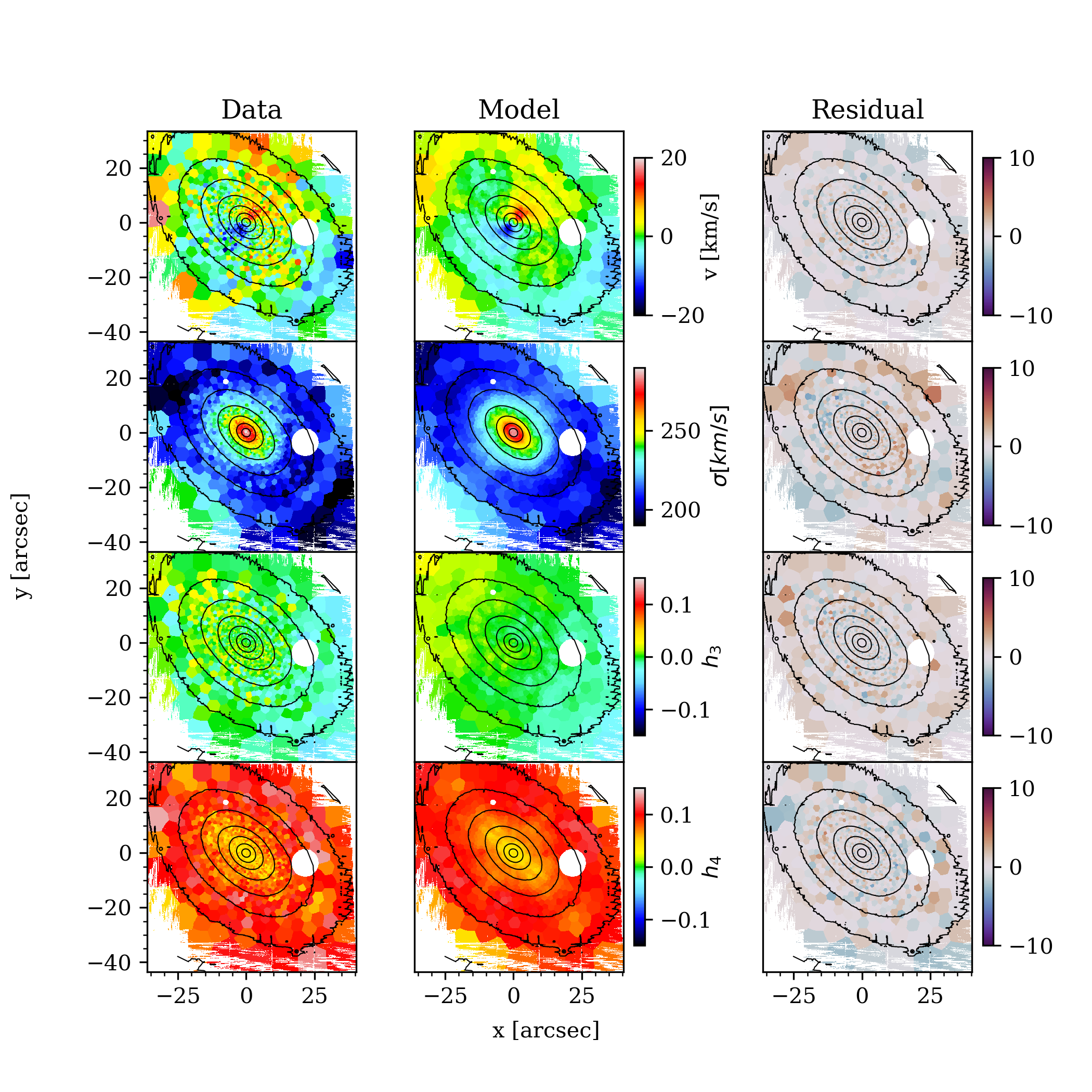}}
    \caption{Observed and modelled stellar kinematic maps for NGC3923. Left panel: SINFONI stellar kinematics and the corresponding DYNAMITE models. From top to bottom, the rows show the mean velocity, velocity dispersion, and the $h_{3}$ and $h_{4}$ Gauss-Hermite higher-order velocity moments. In both panels, the first column shows the observed kinematics from SINFONI (left panel) and MUSE (right panel), the second column shows the corresponding modelled kinematics, and the third column shows the residuals. The residual maps show the difference between the model and the observed kinematics, divided by the observational errors. Overplotted black contours indicate the isophotes.}
    \label{fig:AppendixkinmapNGC3923}
\end{figure*}

\begin{figure*}[h]
    \centering
    \subfigure{
        \includegraphics[width=0.51\textwidth]{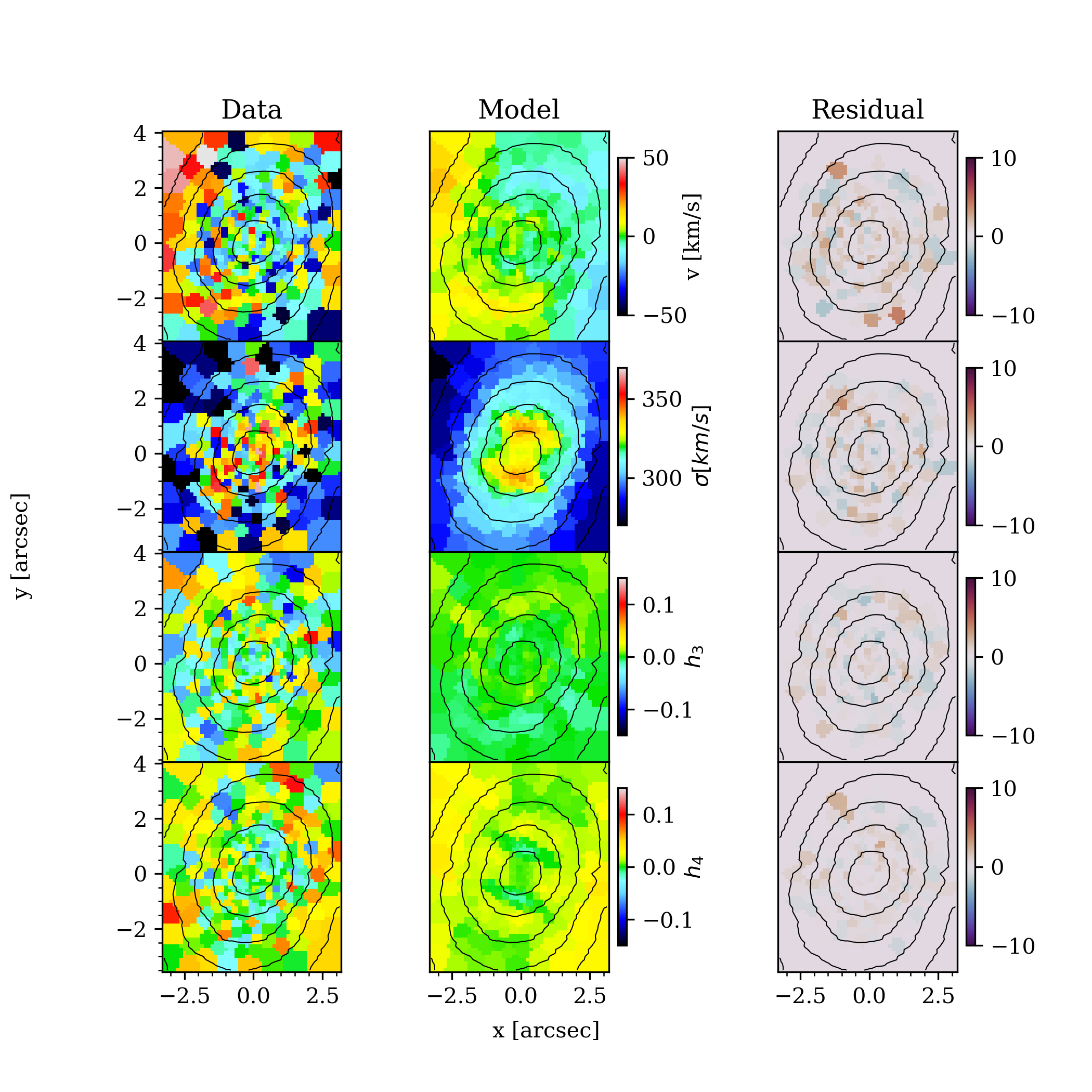}
        \includegraphics[width=0.51\textwidth]{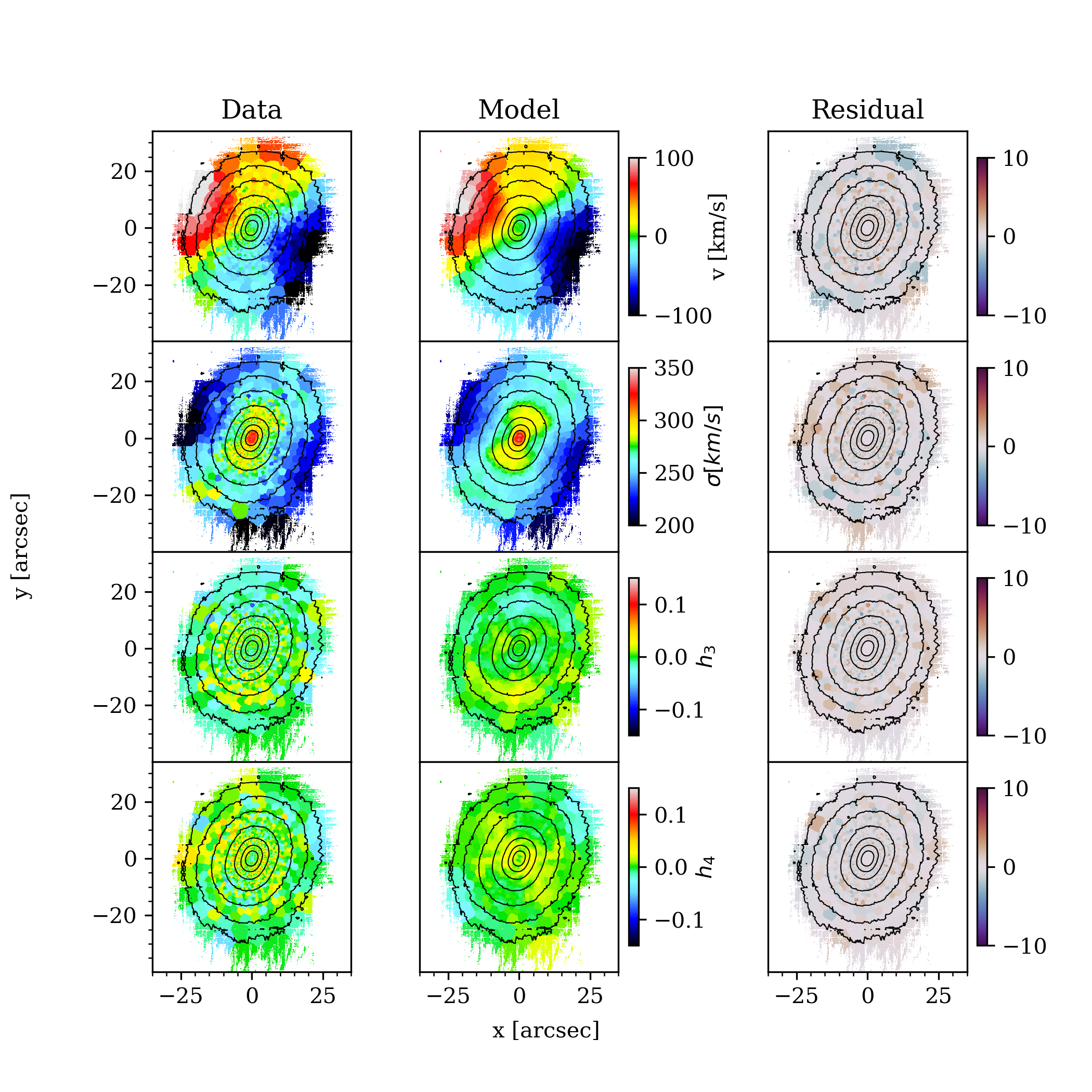}}
    \caption{Same as Figure \ref{fig:AppendixkinmapNGC3923} but for NGC4261}
    \label{fig:AppendixkinmapNGC4261}
\end{figure*}

\begin{figure*}[h]
    \centering
    \subfigure{
        \includegraphics[width=0.51\textwidth]{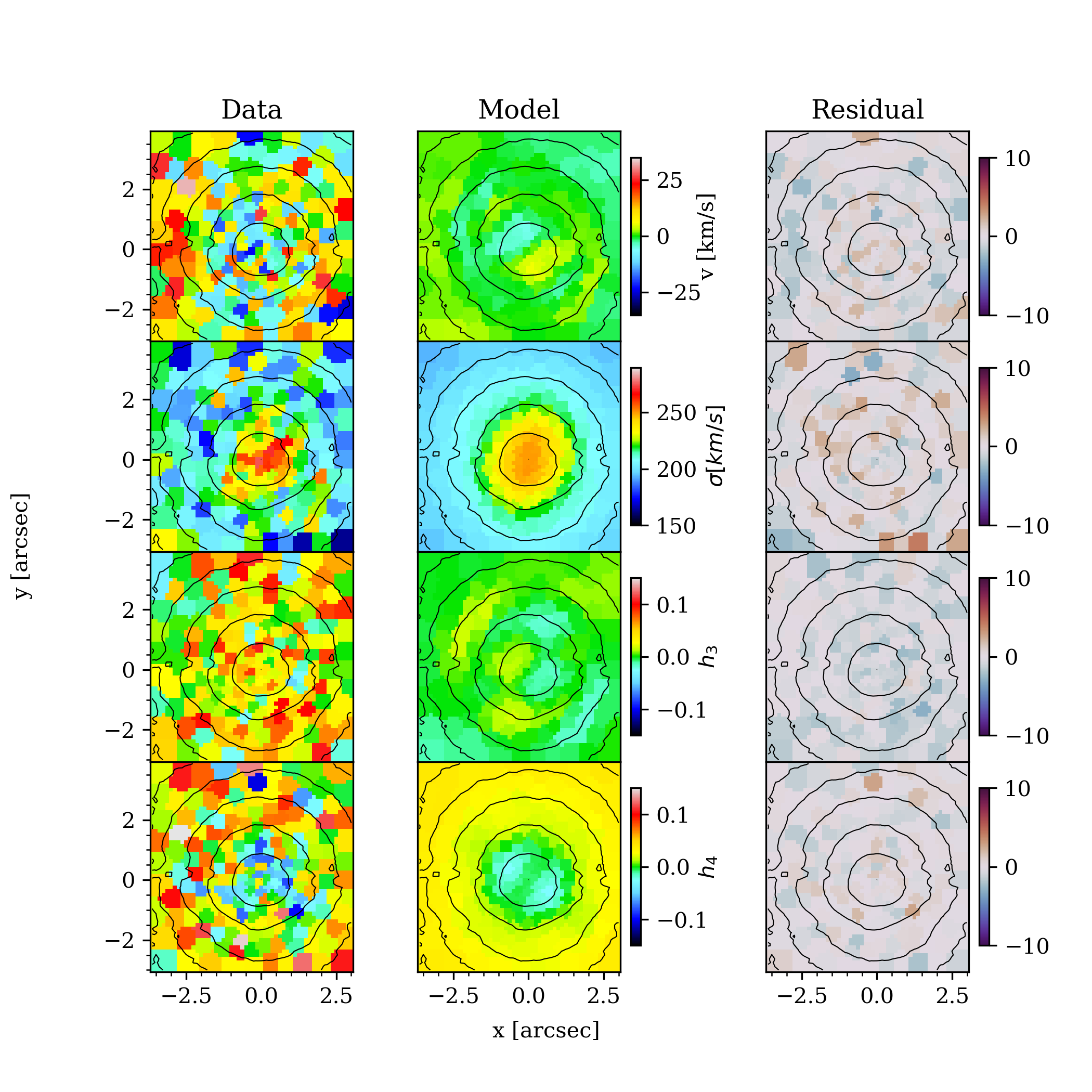}
        \includegraphics[width=0.51\textwidth]{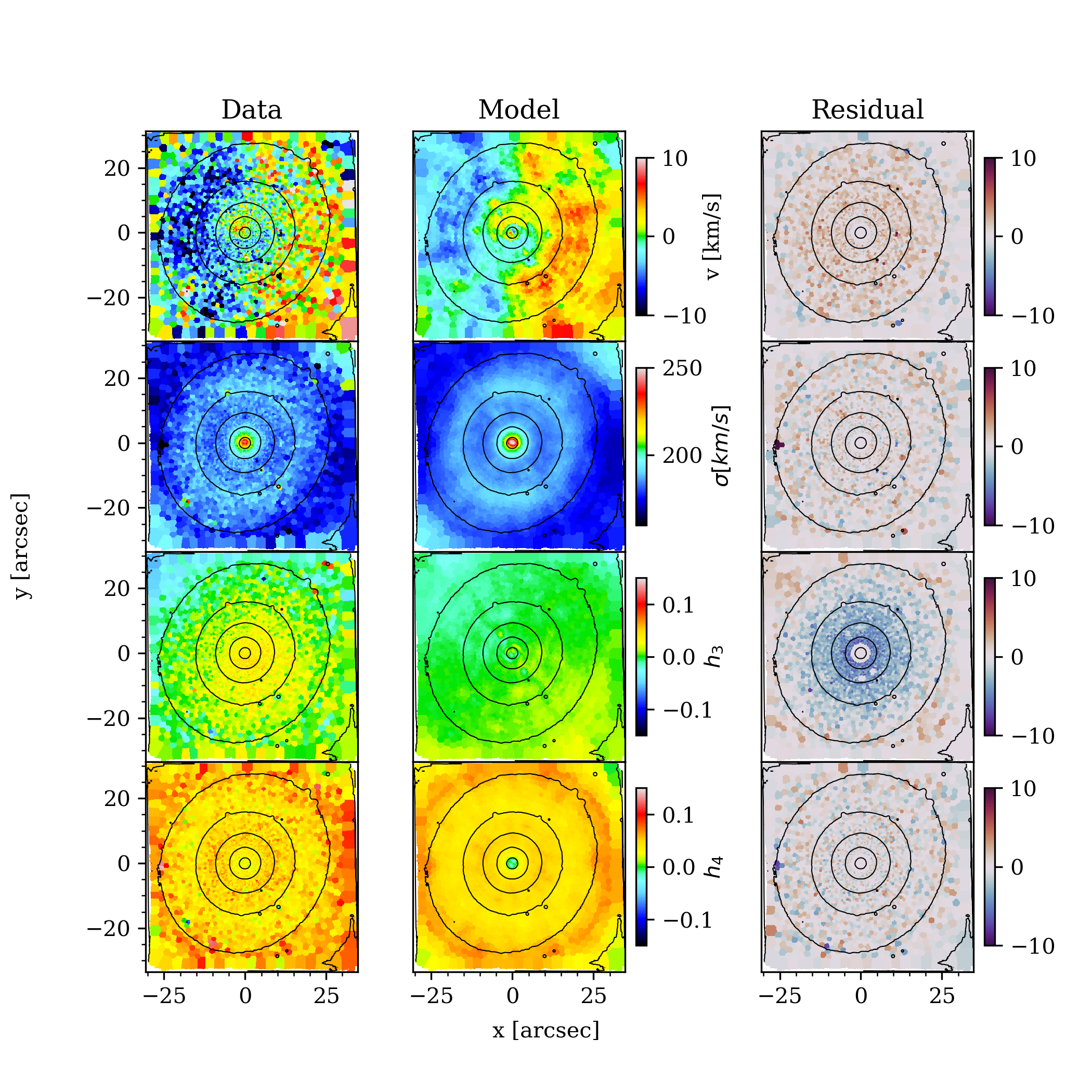}}
    \caption{Same as Figure \ref{fig:AppendixkinmapNGC3923} but for NGC4636}
    \label{fig:AppendixkinmapNGC4636}
\end{figure*}

\begin{figure*}[h]
    \centering
    \subfigure{
        \includegraphics[width=0.50\textwidth]{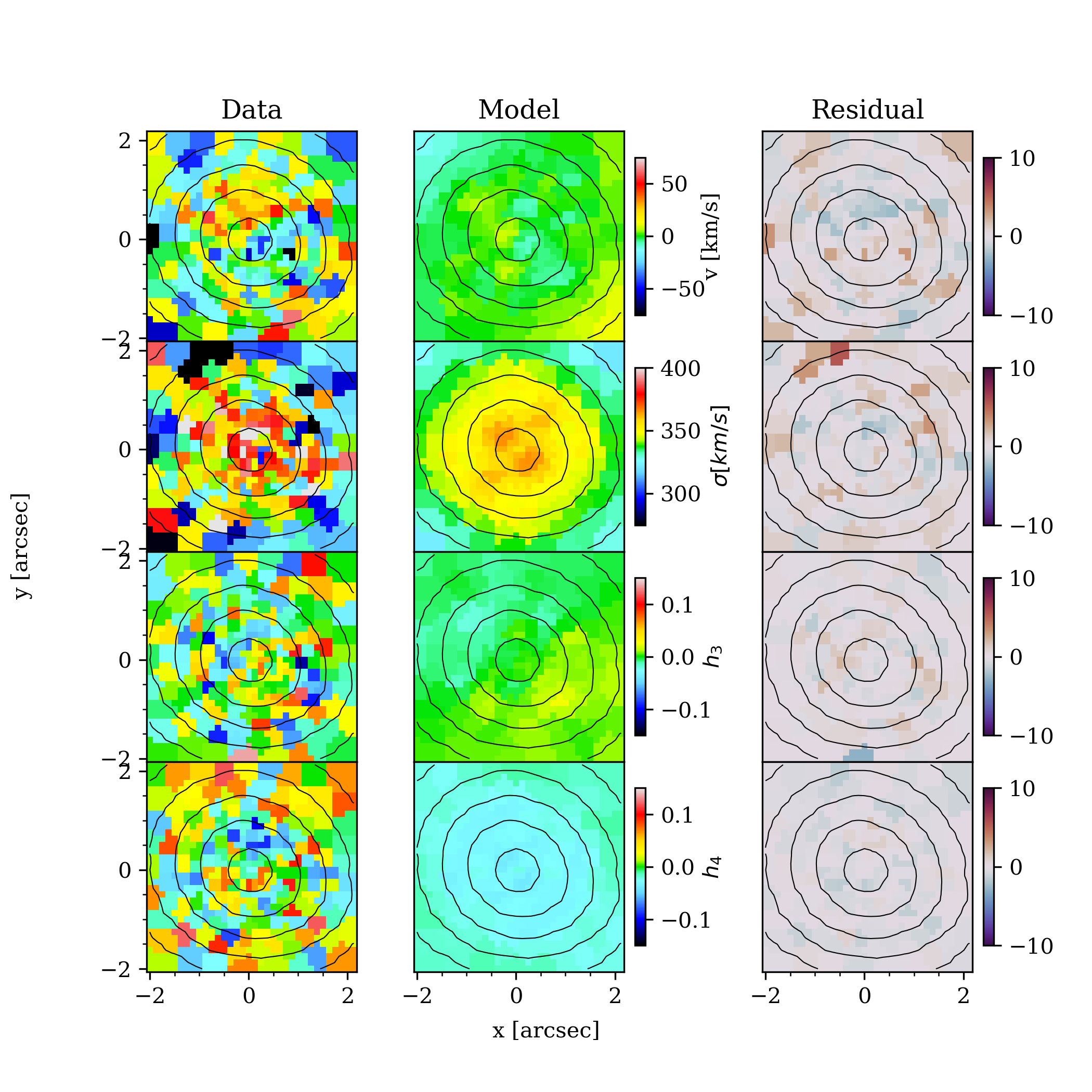}
        \includegraphics[width=0.50\textwidth]{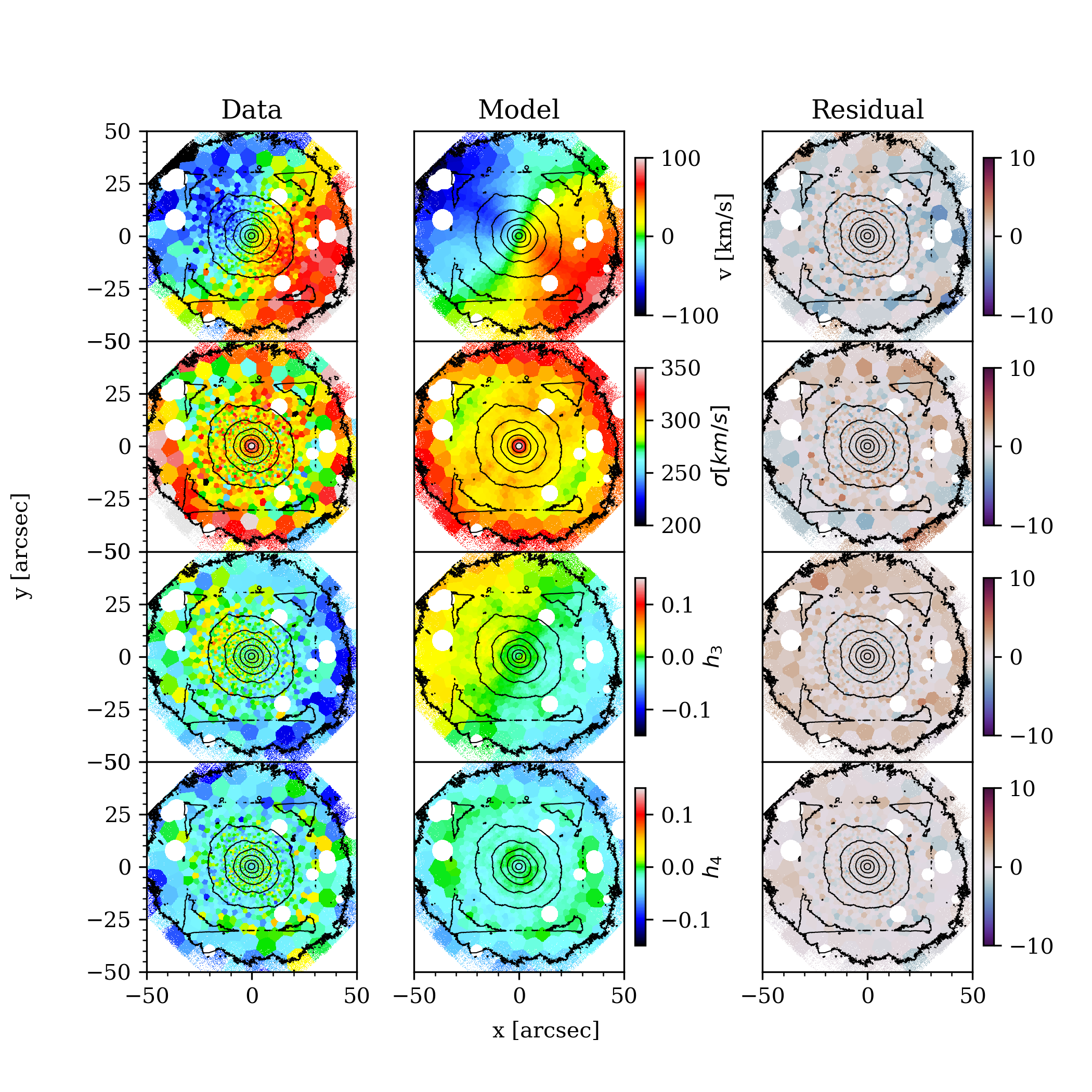}}
    \caption{Same as Figure \ref{fig:AppendixkinmapNGC3923} but for IC4296. The square isophotes visible in the MUSE data (right panel) are artefacts produced by the MUSE pointings.}
    \label{fig:AppendixkinmapIC4296}
\end{figure*}

\begin{figure*}[h]
    \centering
    \subfigure{
        \includegraphics[width=0.50\textwidth]{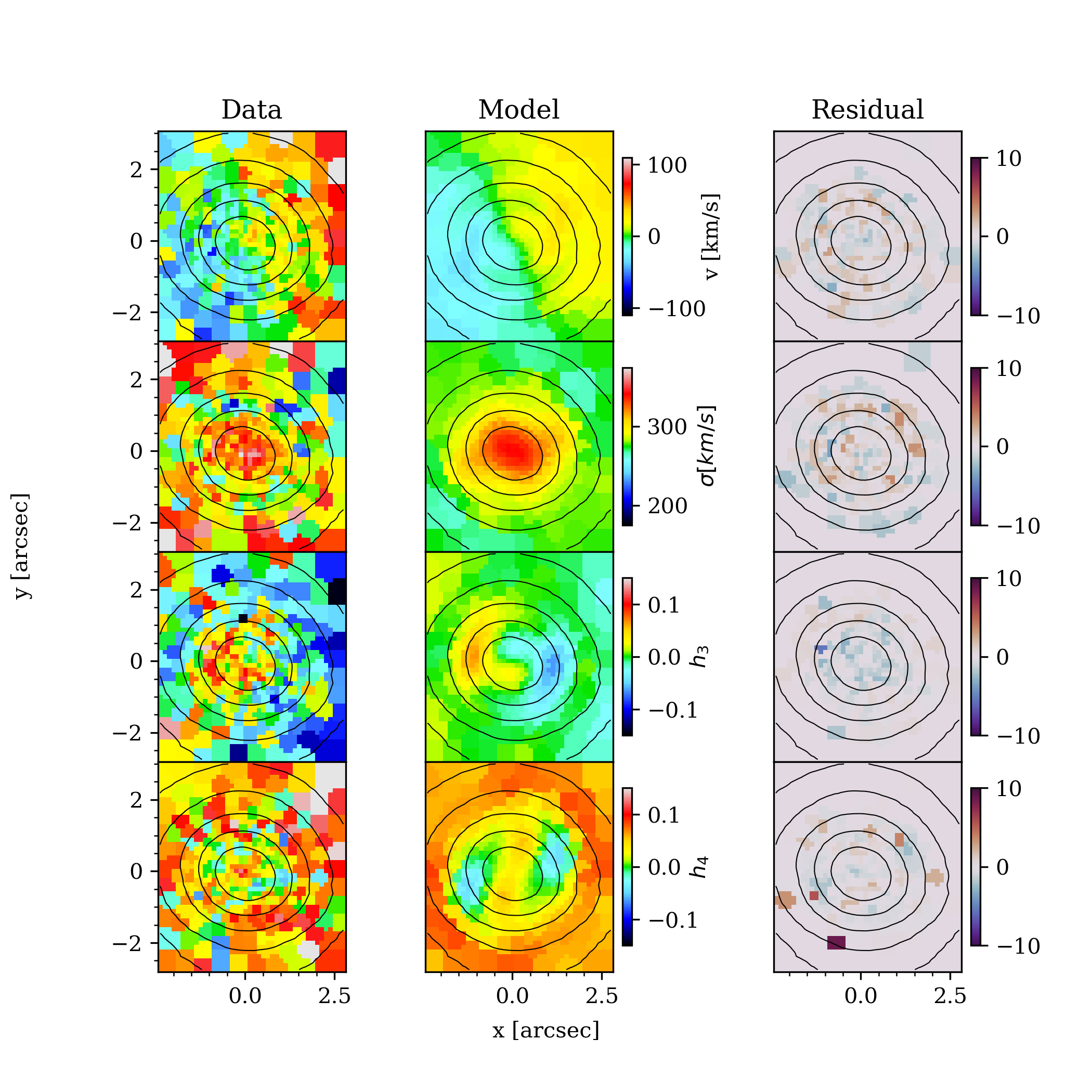}
        \includegraphics[width=0.50\textwidth]{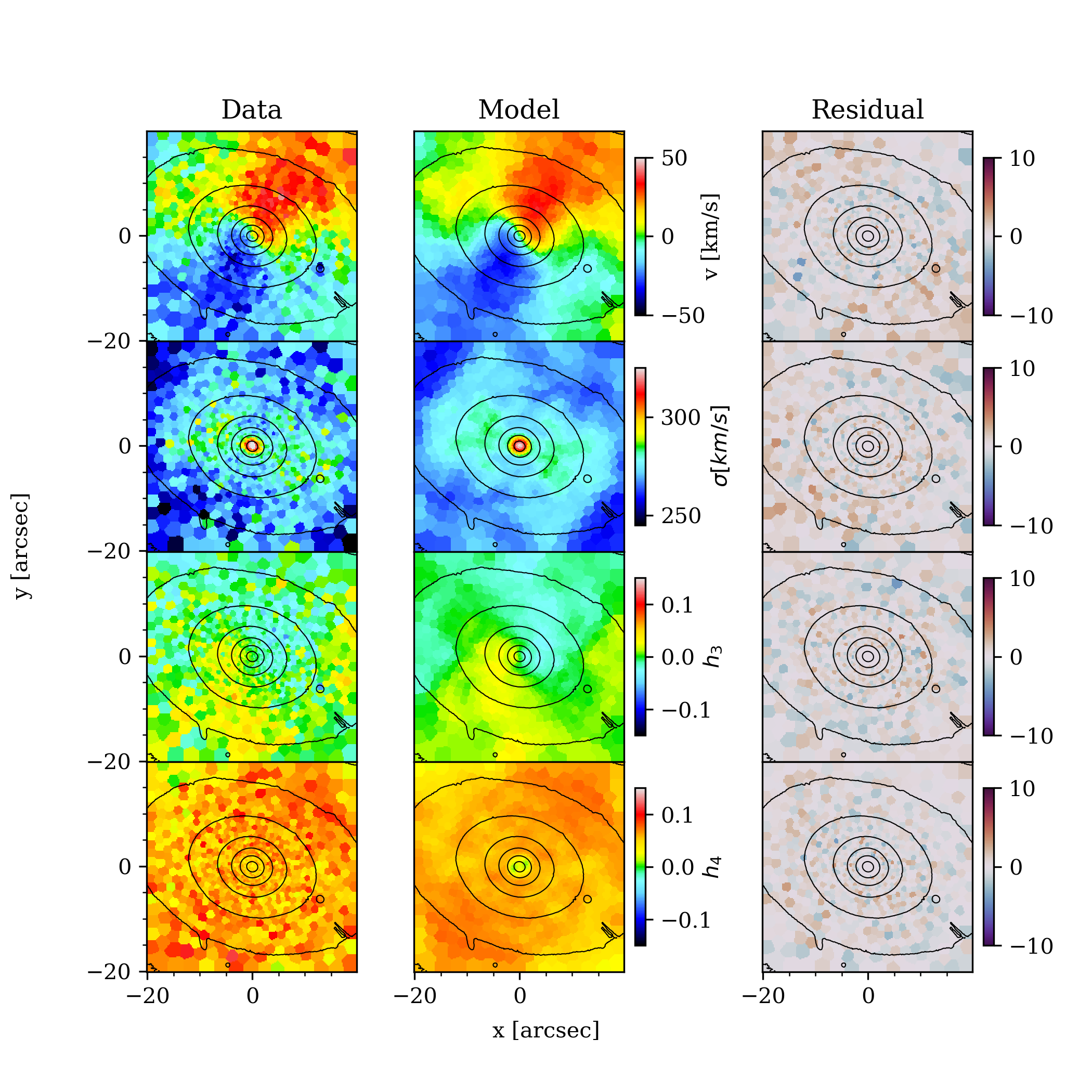}}
    \caption{Same as Figure \ref{fig:AppendixkinmapNGC3923} but for IC4329}
    \label{fig:AppendixkinmapIC4329}
\end{figure*}

\section{Enclosed mass profile} 

\begin{figure*}
\centering
        \includegraphics[width=1.0\linewidth]{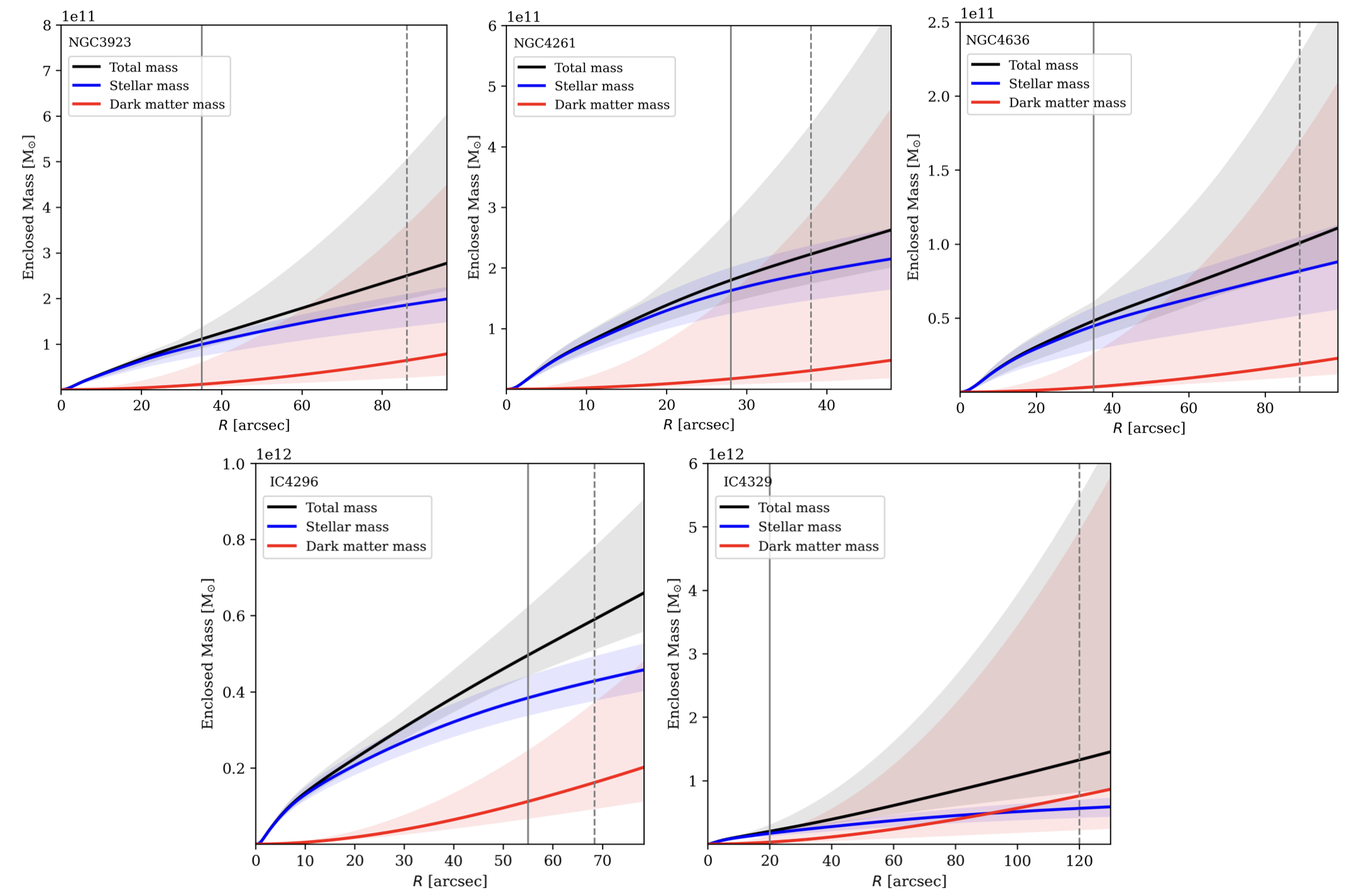}
    \caption{Enclosed mass profile of the five galaxies. Blue and red colors mark the stellar and DM components, while black indicates the total enclosed mass. The shaded regions indicate the 3$\sigma$ uncertainties on the mass profile. The continuous grey and dashed vertical lines show the radial coverage of the MUSE data and the effective radii ($\rm R_{eff}$) of the galaxies, respectively.}
    \label{fig:massprofile}
\end{figure*}

\section{Surface brightness profile and Core-radius of NGC3923} 

\begin{figure}
        \includegraphics[width=1.0\linewidth]{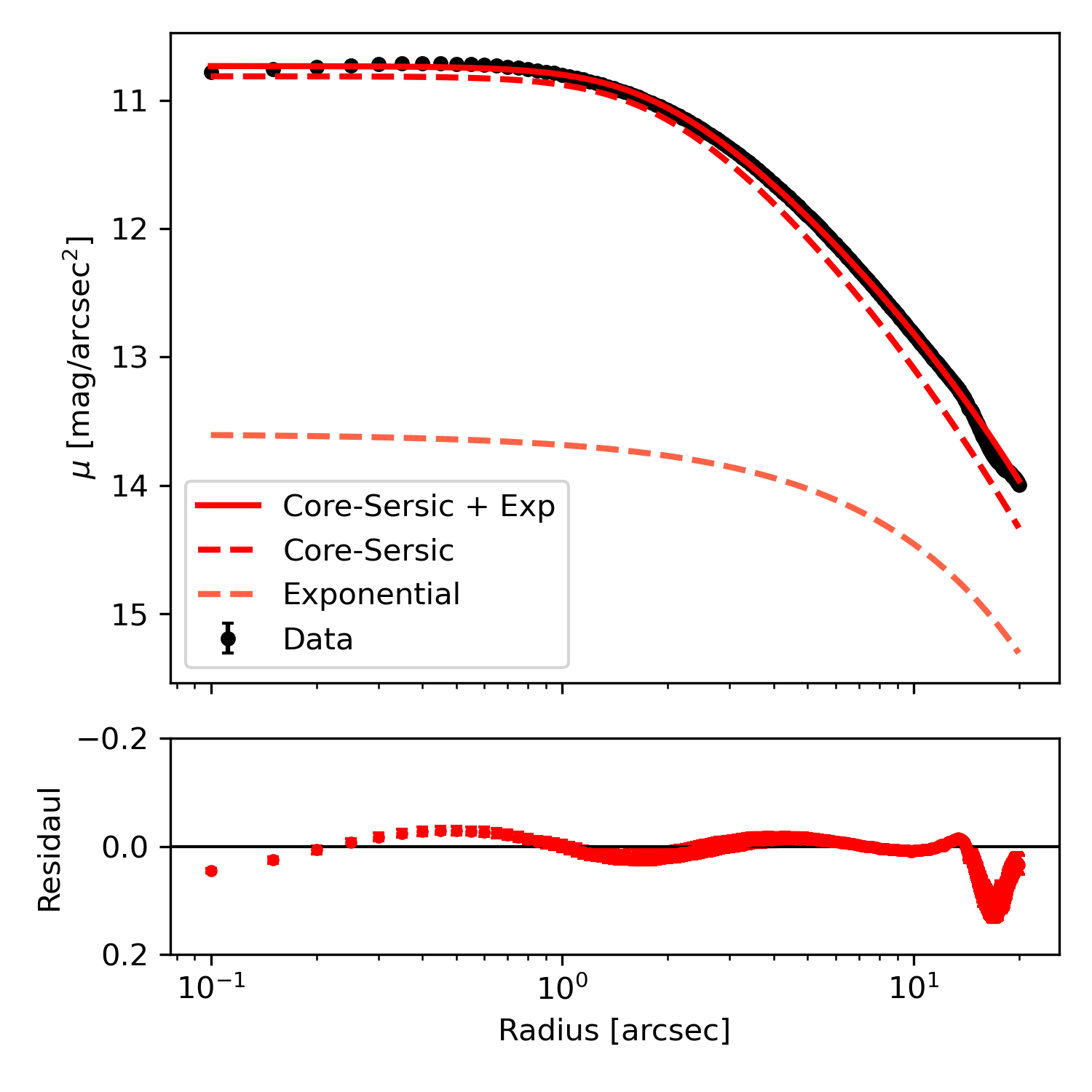}
    \caption{Fit to the surface-brightness profile of NGC3923 obtained from its HST-WFC3/F814W image. Black data points show the extracted 1D surface-brightness profile of NGC3923 along its semi-major axis. The red curve shows the two-component fit, including a core-S'{e}rsic profile as the inner component and an exponential profile as the outer component. The dashed red and light-red curves show the individual contributions of the core-S'{e}rsic and exponential components, respectively, as marked in the plot legend. The lower panel shows the surface-brightness residuals between the model and the data.}
    \label{fig:Appendix_NGC3923_rb}
\end{figure}

\end{appendix}

\end{document}

%% file: All_parameters_appendix.tex
\begin{table*}
\caption{Results of the first step of the Schwarzschild modelling.}
\label{Tab1:results}
\centering
\setlength{\tabcolsep}{3pt}
\renewcommand{\arraystretch}{1.4}

\resizebox{\textwidth}{!}{
\begin{tabular}{lcccccccc}
\toprule

\textbf{Galaxy} &
$\log_{10}(M_{\rm BH}/M_{\odot})$ & $S$ & $p$ & $q$ &
$u$ &
$\theta$ &
$\phi$ &
$\psi$ \\
& 
& $(M_{\rm dyn}/L)/(M_{*}/L)$
&
&
&
&
deg &
deg &
deg \\
\midrule
\textbf{NGC3923} & $8.00 \pm 0.50$ & $1.40 \pm 0.18$ & $0.76 \pm 0.06$ & $0.58 \pm 0.03$ & $0.99 \pm 0.04$ & $77.55 \pm 10.38$ & $75.69 \pm 13.81$ & $91.99 \pm 1.63$ \\
\textbf{NGC4261} & $6.69 \pm 1.50$ & $0.85 \pm 0.13$ & $0.82 \pm 0.05$ & $0.68 \pm 0.03$ & $0.97 \pm 0.04$ & $70.29 \pm 14.63$ & $66.29 \pm 20.71$ & $95.07 \pm 4.30$ \\
\textbf{NGC4636} & $6.50 \pm 0.50$ & $1.07 \pm 0.18$ & $0.85 \pm 0.06$ & $0.62 \pm 0.13$ & $0.93 \pm 0.09$ & $68.53 \pm 12.92$ & $42.53 \pm 21.76$ & $97.11 \pm 5.66$ \\
\textbf{IC4296}   & $7.05 \pm 1.07$ & $1.02 \pm 0.09$ & $0.96 \pm 0.02$ & $0.89 \pm 0.01$ & $1.00 \pm 0.01$ & $68.29 \pm 7.36$ & $87.02 \pm 16.12$ & $90.45 \pm 1.64$ \\
\textbf{IC4329}   & $9.20 \pm 0.20$ & $1.48 \pm 0.19$ & $0.69 \pm 0.03$ & $0.46 \pm 0.05$ & $0.90 \pm 0.04$ & $61.37 \pm 9.98$ & $49.43 \pm 7.29$ & $104.29 \pm 3.84$ \\
\bottomrule
\end{tabular}}

\vspace{0.1cm}

\begin{tabular}{lcccc}
\toprule
\textbf{Galaxy} &
$\rm R_{s}$ &
$\log_{10}(\rm M_{*}/M_{\odot})$ &
$\log_{10} f_{\rm DM}$ &
$\chi^{2}/{\rm DOF}$ \\
&
kpc
&
$M_{200}/M_{*}$ &
\\
\midrule
\textbf{NGC3923}  & 63.02 & 11.60 & $1.31 \pm 0.88$ & $1.27$ \\
\textbf{NGC4261}  & 75.45 & 11.61 & $1.48 \pm 1.94$ & $1.16$ \\
\textbf{NGC4636}  & 25.62 & 11.20 & $0.81 \pm 0.88$ & $3.42$ \\
\textbf{IC4296}   & 51.59 & 11.99 & $0.72 \pm 0.62$ & $1.55$ \\
\textbf{IC4329}   & 85.92 & 11.91 & $1.31 \pm 0.91$ & $1.13$ \\
\bottomrule
\end{tabular}

\tablefoot{Best-fit model parameters obtained by simultaneously fitting the SINFONI and large-scale MUSE kinematics while allowing all model parameters to vary freely. The BH masses obtained in this step are intermediate estimates and are not the final adopted values. The upper table gives the best-fit black-hole mass, stellar mass-to-light scaling factor $S=(M_{\rm dyn}/L)/(M_{*}/L)$, intrinsic shape parameters $p$ (intrinsic medium-to-major axis ratio), $q$ (intrinsic minor-to-major axis ratio), and $u$ (ratio of the projected to intrinsic major axis), and viewing angles $\theta$, $\phi$, and $\psi$. The lower table gives the total stellar mass, obtained by summing the stellar MGE components constructed using the radially varying $\rm M_{*}/L$ profile and applying the best-fitting scaling factor $\rm S$, the NFW halo scale radius $\rm R_{\rm s}$, $\rm f_{\rm DM}=M_{200}/M_{*}$, the dark matter fraction at $r_{200}$, and the reduced chi-square of the best-fitting model, $\chi^{2}/{\rm DOF}$. The quoted uncertainties correspond to projected $1\sigma$ confidence intervals from the explored model grid.}

\end{table*}

%% file: BHmass.tex
\begin{table}
\caption{Best-fitting black hole masses and scaling factor from the Schwarzschild models.}

\label{Tab2:results}
\centering
\setlength{\tabcolsep}{3pt}
\renewcommand{\arraystretch}{1.6} % Default value: 1
  \begin{tabular}{c|lll}  % Ensure column count matches
    \toprule
    \textbf{Galaxy} & $\rm M_{BH}$ & $\rm S$ & $\rm \chi^{2}/DOF$  \\
        & ($10^{8} \ \rm M_{\odot}$) &  \ $\rm (M_{dyn}/L)/(M_{*}/L)$ & \\  % Units row
    (1) & (2) & (3) & (4)          \\  % Column number reference
    \midrule
    % Data rows go here
    \textbf{NGC3923} & $1.99^{+6.76}_{-1.05}$ & $1.30^{+0.15}_{-0.21} $ & 1.28\\
    \textbf{NGC4261} & $<$ 28.27  &  $0.75^{+0.23}_{-0.23}$  & 1.09 \\
    \textbf{NGC4636} & $<$ 12.40  &  $0.88^{+0.37}_{-0.18}$ & 3.42\\
    \textbf{IC4296}  & $<$ 33.57 & $1.11^{+0.10}_{-0.10} $ & 1.58  \\
    \textbf{IC4329}  & $21.78^{+13.58}_{-11.32}$ & $1.44^{+0.30}_{-0.16} $ & 1.12  \\
    \hline
  \end{tabular}
 \tablefoot{This table summarizes the BH mass estimates from the second step of Schwarzschild modelling using the MUSE and SINFONI kinematics. In this step, only the BH mass and scaling factor S are varied, while all other parameters are fixed to their best-fitting values from the first modelling step. Column 1: Name of the modelled galaxy. Columns 2 and 3: Best-fit values of $\rm M_{BH}$ and scaling factor S, with uncertainties calculated at the 3$\sigma$ confidence level. Column 4: $\rm \chi^{2}$ value of the best-fitting model divided by the degrees of freedom.}   
\end{table}